\documentclass[journal]{IEEEtran}
\IEEEoverridecommandlockouts

\usepackage{cite}
\usepackage{amsmath,amssymb,amsfonts}
\usepackage{algorithm}
\usepackage{algpseudocode}
\usepackage{graphicx}
\usepackage{caption}
\usepackage{subcaption}
\usepackage[compact]{titlesec}  
\titlespacing{\section}{2pt}{2pt}{2pt}
\usepackage{textcomp}
\usepackage{xcolor}
\usepackage{comment}
\usepackage{hyperref}
\usepackage{graphicx}
\def\BibTeX{{\rm B\kern-.05em{\sc i\kern-.025em b}\kern-.08em
    T\kern-.1667em\lower.7ex\hbox{E}\kern-.125emX}}
\usepackage{bm}
\usepackage{cases}
\usepackage{array}
\usepackage{booktabs}
\usepackage{pifont}
\newcommand{\cmark}{\ding{51}}
\newcommand{\xmark}{\ding{55}}
\usepackage[font=footnotesize]{caption}
\usepackage{makecell}

\begin{document}

\title{Accelerating Distributed Variational Quantum Computing over Wireless Networks 
with Incentive Mechanism Design}

\author{\IEEEauthorblockN{Shaba Shaon, Dinh C. Nguyen,~\textit{Member, IEEE}, and Octavia Dobre,~\textit{Fellow, IEEE}} 
\vspace{-15pt}
\thanks{Shaba Shaon and Dinh C. Nguyen are with ECE Department,  University of Alabama in Huntsville, Huntsville, AL 35899, USA, emails: (ss0670@uah.edu, dinh.nguyen@uah.edu).}
\thanks{Octavia Dobre is with the Faculty of Engineering and Applied Science, Memorial University, Canada (e-mail: odobre@mun.ca).}
\thanks{*This work was supported in part by the U.S. National Science Foundation
(NSF) under Grant SaTC-2513164. The work of Dr. Octavia Dobre was supported in part by the Canada Research Chairs Program CRC-2022-00187 and the Natural Sciences and Engineering Research Council of Canada (NSERC), Discovery Program, Grant RGPIN-2026-04392.}
}

\maketitle

\begin{abstract}
Distributed variational quantum computing (DVQC) appears as the emerging paradigm to address the limitations of near-term quantum devices by decomposing large parameterized quantum circuits (PQCs) into smaller subcircuits executed across multiple quantum processors. In this context, how to incentivize quantum processors to participate actively is crucial for accelerating DVQC, yet remains largely unexplored in the literature. To address this significant gap, we propose a novel two-stage incentive framework for wireless DVQC. In the first stage, we minimize processor-side energy cost via a joint optimization of subcircuit parameters, number of noise measurements, and transmit power, efficiently solved using block coordinate descent (BCD) and successive convex approximation (SCA). In the second stage, we maximize aggregator-side system utility by selecting an optimal subset of processors based on their optimized energy costs derived from the first stage, formulating a non-convex combinatorial selection problem addressed using the alternating direction method of multipliers (ADMM). \textcolor{black}{Extensive numerical experiments on the MNIST and CIFAR-10 datasets demonstrate that our DVQC framework achieves training and testing performance comparable to centralized quantum computing in terms of accuracy and loss, with similar convergence behavior on MNIST and slightly slower initial convergence on CIFAR-10.} Moreover, our optimization approach consistently outperforms baseline schemes in energy efficiency and system utility, revealing a strong synergy between processor-level energy minimization and incentive-driven processor selection, where reduced energy costs directly enhances a processor’s likelihood of selection and contributes to higher overall system utility.

\end{abstract}

\section{Introduction} \label{sectionI}
Variational quantum computing (VQC) has emerged as a powerful hybrid computing paradigm for leveraging the capabilities of near-term quantum devices to tackle complex learning and optimization problems \cite{chen2020variational}. This approach employs parameterized quantum circuits (PQCs), where circuit parameters are iteratively tuned using classical optimization routines. By combining quantum state evolution with classical feedback, variational quantum computing can execute meaningful computations even under limited coherence times, restricted connectivity, and qubit counts of today’s hardware \cite{fellous2023optimizing}.

However, scaling variational quantum computing to solve real-world problems remains challenging due to qubit resource constraints on \textit{near-term} quantum devices. Distributed variational quantum computing (DVQC) offers a promising approach to scaling quantum workloads by distributing computation across a network of quantum processors \cite{he2024distributionally}. However, a key challenge lies in motivating individual processors to participate. Each processor must expend valuable resources, such as energy, communication bandwidth, and the finite duration over which quantum information can be preserved, i.e., quantum coherence time, to perform its assigned quantum operations and report results. Without effective incentives, processors may act selfishly or refuse to collaborate, compromising both the reliability and scalability of the distributed system. \textcolor{black}{Therefore, designing an effective incentive mechanism is essential to ensure cooperative participation of quantum processors while optimizing system level performance. This work aims to improve energy consumption, system utility, and communication performance in DVQC networks under practical constraints such as limited qubit resources, transmission overhead, and processor self-interest.}

\subsection{Related Work}
\textbf{Variational Quantum Computing:} Recently, a diverse range of VQC schemes have been explored. The work in \cite{mastroianni2024variational} investigated how VQC can address the NP-hard problem of optimally assigning and scheduling computations across heterogeneous cloud or edge architectures. The authors in \cite{schillo2025variational} explored how VQC can be used to solve differential equations on near-term quantum devices. In \cite{roosan2025variational} the authors presented a hybrid quantum-classical graph neural network framework that uses VQCs and quantum approximate optimization algorithm to boost molecular classification performance, demonstrating the advantages of quantum computing in complex biochemical tasks. The authors in \cite{huang2022learning} used VQCs enhanced by a meta-learning framework to improve training efficiency and scalability in solving optimization and molecular simulation tasks on near-term quantum devices devices. Besides, the work in \cite{wurtz2021classically, chen2020variational, venkatesh2024qubit, he2024distributionally} leverage VQCs for solving combinatorial optimization, reinforcement learning, image segmentation, and robust parameter training under noise, respectively. The authors in \cite{saghafi2024variational} and \cite{zhang2024hybrid} utilized VQCs for solving power system analytics via quantum linear solvers and for optimizing beamforming in wireless communications through hybrid quantum-classical neural networks, respectively. \textcolor{black}{\textit{Although all these works leverage VQCs to solve diverse application problems, they generally assume the availability of sufficiently large and reliable quantum hardware. In practice, however, near-term quantum devices are severely qubit-limited and constrained by noise, decoherence, and imperfect connectivity, which often prevents the execution of large monolithic PQCs and, in some cases, even moderate-scale circuits. Consequently, these studies do not consider distributed quantum computing frameworks to overcome such hardware limitations, nor do they incorporate incentive mechanisms to motivate participation of independently operated processors in a distributed quantum environment.}}

\begin{table}[t]
\centering
\caption{\textcolor{black}{Comparison with related work}}
\label{tablerelatedwork}
\setlength{\tabcolsep}{3pt}
\renewcommand{\arraystretch}{1.0}
{\color{black}
\begin{tabular}{|>{\centering\arraybackslash}p{2.5cm}|
                >{\centering\arraybackslash}c|
                >{\centering\arraybackslash}c|
                >{\centering\arraybackslash}c|
                >{\centering\arraybackslash}c|
                >{\centering\arraybackslash}c|
                >{\centering\arraybackslash}c|}
\hline
Features & 
\begin{tabular}{@{}c@{}} 
\cite{chen2020variational}, \\
\cite{wurtz2021classically, mastroianni2024variational, schillo2025variational, roosan2025variational, huang2022learning, venkatesh2024qubit, he2024distributionally} 
\end{tabular}
& \cite{saghafi2024variational} &\cite{zhang2024hybrid} & \cite{tang2021cutqc, brandhofer2023optimal, peng2020simulating} & \cite{kan2024scalable, pawar2024qrcc, shehzad2024automated}  & 
\begin{tabular}{@{}c@{}}
Our \\
work
\end{tabular} \\
\hline
General VQC 
& \cmark & \cmark & \cmark & \xmark & \xmark & \cmark \\
\hline
VQC for wireless or power systems
& \xmark & \cmark & \cmark & \xmark & \xmark & \cmark \\
\hline
DVQC 
& \xmark & \xmark & \xmark & \cmark & \cmark & \cmark \\
\hline
Circuit cutting 
& \xmark & \xmark & \xmark & \cmark & \cmark & \cmark \\
\hline
Wireless communication modeling 
& \xmark & \xmark & \cmark & \xmark & \xmark & \cmark \\
\hline
NOMA-based Uplink
& \xmark & \xmark & \xmark & \xmark & \xmark & \cmark \\
\hline
Processor energy optimization
& \xmark & \xmark & \xmark & \xmark & \xmark & \cmark \\
\hline
System utility optimization 
& \xmark & \xmark & \xmark & \xmark & \xmark & \cmark \\
\hline
Incentive mechanism design for processor
& \xmark & \xmark & \xmark & \xmark & \xmark & \cmark \\
\hline
Optimization based on BCD, SCA, ADMM
& \xmark & \xmark & \xmark & \xmark & \xmark & \cmark \\
\hline
\end{tabular}}
\end{table}
\textbf{Distributed VQC over Networks:} 
A few works have addressed qubit limitations by employing circuit cutting techniques to partition large quantum circuits across networked quantum processors. In \cite{tang2021cutqc}, the authors proposed a hybrid quantum-classical framework that enables execution of large quantum circuits by partitioning them into smaller subcircuits runnable on limited near-term quantum devices, followed by classical postprocessing to reconstruct the full output. The work in \cite{brandhofer2023optimal} presented a quantum circuit partitioning method using quantum circuit knitting to minimize qubit requirements and reduce runtime, enabling efficient execution on near-term quantum hardware. In \cite{peng2020simulating}, a cluster-based simulation scheme was introduced that enables quantum circuits with limited inter-cluster communication to be efficiently executed on small quantum devices, reducing qubit requirements for tasks such as variational eigensolvers. The works in \cite{kan2024scalable, pawar2024qrcc, shehzad2024automated} also leveraged circuit cutting techniques to facilitate efficient distributed quantum computing across a range of application domains. \textcolor{black}{\textit{However, these studies implicitly assume that distributed quantum processors are readily available and willing to participate in cooperative execution. In practical, market-driven environments, quantum processors are often owned by independent entities and may not participate without proper incentives, making incentive design a crucial requirement that is largely missing in the existing literature. Moreover, prior works do not consider energy minimization of quantum computation and measurement transmission, nor do they optimize system utility in a distributed setting. Importantly, none of the existing studies model a realistic communication uplink, such as power-domain NOMA, or integrate networking considerations into the DVQC framework. Since transmitting measurement outcomes to the aggregator is a core and recurring operation in DVQC, incorporating wireless communication modeling from a networking perspective is essential for practical deployment, yet remains unaddressed by existing works.}} \textcolor{black}{A comprehensive comparison between the proposed framework and existing studies is provided in Table~\ref{tablerelatedwork} to highlight the key differences and the unique contributions of this work.}

\subsection{Our Contributions}
\textcolor{black}{As discussed above, despite significant research efforts on VQCs and circuit cutting techniques, existing works do not jointly address several practical challenges. Current near-term quantum hardware is severely qubit-limited and noise-prone, making large monolithic PQCs infeasible. Moreover, quantum processors are energy-constrained devices, and repeated circuit execution and measurement transmission incur substantial energy overhead. In realistic market-driven environments, quantum processors are owned by independent entities that act in their self-interest and may not participate in distributed computation unless adequately compensated for their energy expenditure. Furthermore, since DVQC critically relies on transmitting measurement outcomes to an aggregator, it is essential to model a realistic wireless uplink with interference consideration, such as power-domain NOMA. Therefore, jointly addressing qubit limitations, incentive-driven participation, system utility maximization, energy efficiency, and communication-aware design is necessary for a holistic and practically deployable DVQC framework. \textcolor{black}{While existing works have addressed several individual aspects of VQC, circuit cutting, and distributed quantum computing, no prior work jointly integrates distributed VQC, processor's energy optimization, incentive-aware processor selection, system utility maximization, and communication-aware wireless networking into a unified framework suitable for practical deployment.} Motivated by these gaps, we propose a unified energy-, communication-, and incentive-aware DVQC framework tailored for distributed quantum processor networks. \textcolor{black}{Our novelty lies in jointly integrating circuit cutting based distributed VQC, processor-side energy optimization, incentive-compatible processor selection, system utility maximization, and a practical NOMA-based uplink model with interference consideration into a single coherent framework.} Specifically, our contributions are summarized as follows:
\begin{itemize}
    \item We develop a novel DVQC framework in which large PQCs are decomposed into smaller subcircuits and executed in parallel across multiple low-qubit and independent quantum processors. The processors communicate measurement outcomes to a central aggregator through a power-domain NOMA uplink with interference modeling and successive interference cancellation (SIC), enabling practical wireless-assisted DVQC deployment (Section~\ref{sectioniII}).
    \item We introduce a novel two-stage incentive framework for DVQC, where quantum processors, owned by independent entities, act in their self-interest and participate if only their energy costs are properly compensated (Section~\ref{sectioniII}). In the first stage, each processor solves a non-convex energy minimization problem. This problem is efficiently solved via block coordinate descent (BCD) and successive convex approximation (SCA). Each processor reports its minimized energy cost to the aggregator. Based on these optimized energy costs, the aggregator then solves a system utility maximization problem to select an optimal set of processors under payment budget and qubit coverage constraints. This selection, formulated as a non-convex combinatorial NP-hard problem, is addressed using alternating direction method of multipliers (ADMM) (Section~\ref{sectionIII}). 
    \item We conduct extensive numerical experiments to evaluate the learning performance, convergence behavior, and system-level efficiency of the proposed DVQC framework. \textcolor{black}{Results on the MNIST and CIFAR-10 datasets show that DVQC achieves training and testing performance comparable to centralized quantum computing in terms of accuracy and loss, with similar convergence behavior on MNIST and slightly slower initial convergence on CIFAR-10.} We further evaluate the proposed energy minimization and system utility maximization frameworks, demonstrating significant gains over baseline strategies. In addition, we compare the proposed NOMA-based uplink with a TDMA scheme, showing that NOMA provides superior energy efficiency and lower transmission latency due to simultaneous measurement outcome transmission. These results confirm the practicality and effectiveness of the proposed framework, particularly in real-world scenarios (Section~\ref{sectionIV}).
\end{itemize}}

\noindent
\textcolor{black}{The principal symbols and notations used throughout this work are summarized in Table~\ref{tablenotations}.}

\begin{figure}[t!]
    \centering
    \includegraphics[width=1\linewidth]{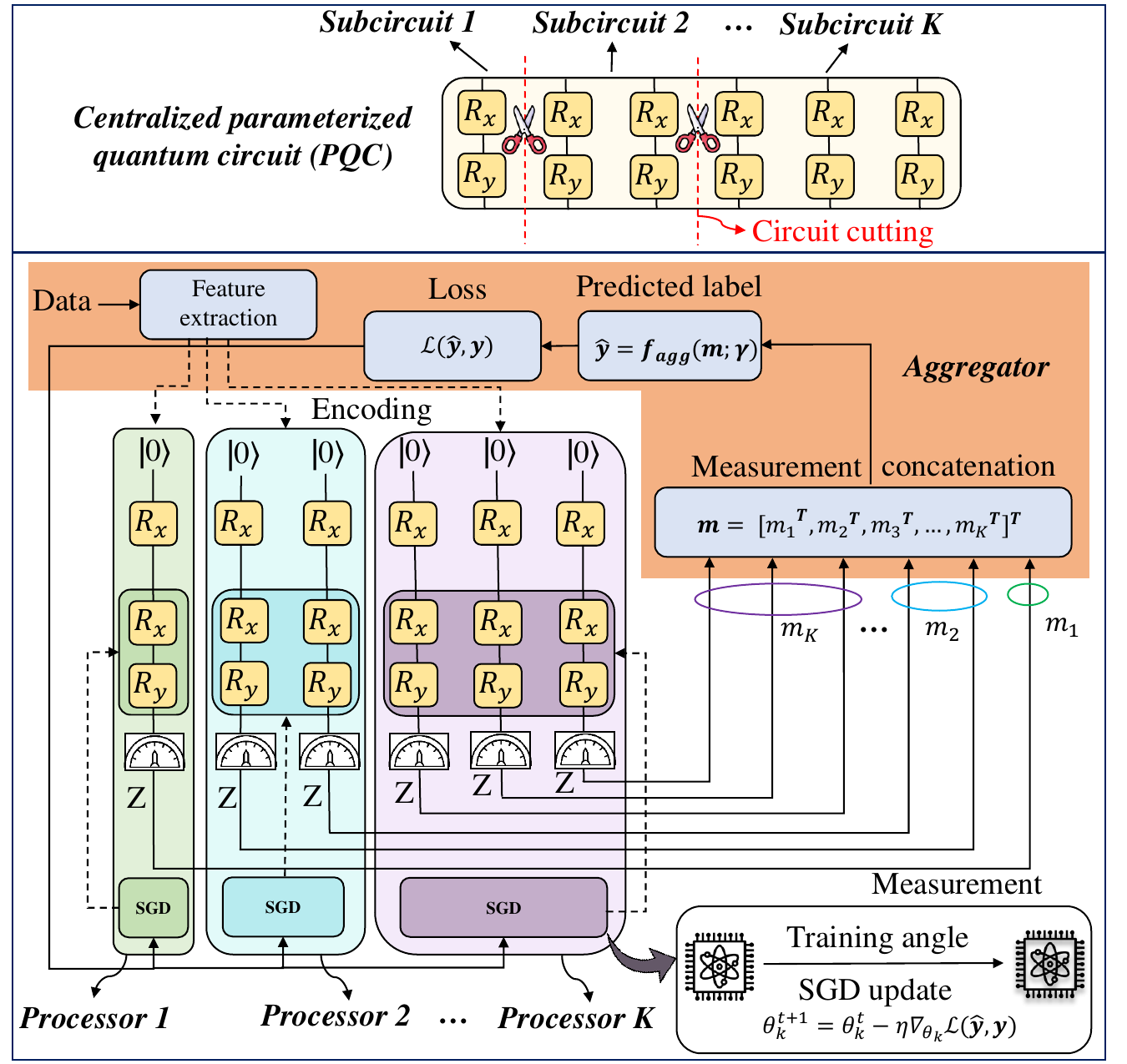} 
    \caption{DVQC framework where large PQCs are decomposed into smaller subcircuits that are executed in parallel across independent quantum processors, while a central aggregator orchestrates the process.}
    \label{fig:overview}  
\end{figure}

\begin{table}[h]
\centering
\caption{\textcolor{black}{List of key notations}}
\label{tablenotations}
\renewcommand{\arraystretch}{1.1}
\setlength{\tabcolsep}{4pt}
\footnotesize
{\color{black}
\begin{tabular}{|c|p{2.6cm}|c|p{2.6cm}|}
\hline
\textbf{Notation} & \textbf{Definition} &
\textbf{Notation} & \textbf{Definition} \\
\hline

$K$ & Number of quantum processors &
$\mathcal{S}$ & Optimal subset of processors \\
\hline

$\mathbf{f}_{k}$ & Input features &
$\mathbf{c}_{k}^{(t)}$ & Measurement output \\
\hline

$x_{k}$ & Selection variable &
$\theta$ & Trainable parameters \\
\hline

$\hat{O}_{k}$ & Observable &
$\phi_{k}^{t}$ & Number of shots \\
\hline

$f_{\text{agg}}(\cdot;\gamma)$ & Post-processing function &
$\hat{y}$ & Final prediction \\
\hline

$y$ & ground-truth label &
$\mathcal{L}(\hat{y}, y)$ & Training loss \\
\hline

$T$ & DVQC iterations &
$E_{k}^{\text{quant}}$ & Quantum computing energy \\
\hline

$E_{k}^{\text{trans}}$ & Measurement transmission energy &
$E_{k}$ & Processor's combined energy \\
\hline

$s_{k}^{(t)}$ & Measurement data size &
$R_{k}^{(t)}$ & Uplink rate \\
\hline

$\gamma_{k}^{(t)}$ & SNR &
$B$ & Bandwidth \\
\hline

$p_{k,\text{tra}}^{(t)}$ & Transmit power &
$\sigma^{2}$ & Gaussian noise $i$ \\
\hline

$A_{k}^{(t)}$ & Signal power &
$B_{k}^{(t)}$ & Interference \\
\hline

$p_{k,1}^{(t)}$ & 1-qubit gate power &
$p_{k,2}^{(t)}$ & 2-qubit gate power \\
\hline

$p_{k,M}^{(t)}$ & Measurement gate power &
$N_{k,1}$ & Average 1-qubit gate count \\
\hline

$N_{k,2}$ & Average 2-qubit gate count &
$N_{k,M}$ & Average measurement gate count \\
\hline

$m_{k}$ & Monetary payment &
$M_{\text{max}}$ & Payment budget \\
\hline

\end{tabular}}
\end{table}

\section{System Model and Problem Formulation} \label{sectioniII}
\subsection{DVQC Learning Model}
\begin{figure}[t!]
    \centering
    \includegraphics[width=1\linewidth]{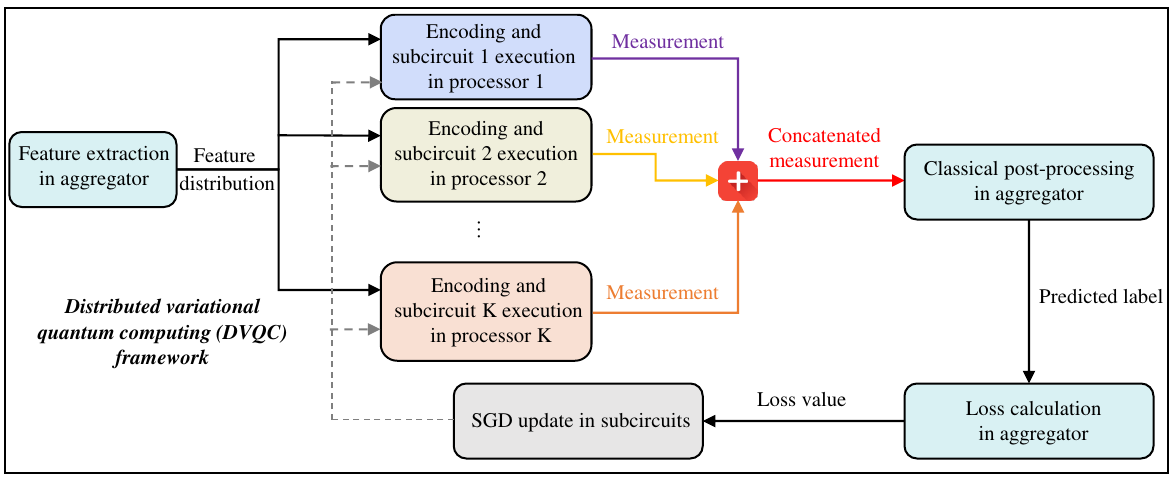} 
    \caption{Proposed DVQC framework with parallel subcircuit execution and central aggregator coordination.}
    \label{fig:overviewblockdiagram}  
\end{figure}

\textcolor{black}{We propose a DVQC framework, as shown in Fig.~\ref{fig:overview}, where a PQC is decomposed into multiple smaller subcircuits leveraging circuit cutting techniques, each executed independently on a dedicated quantum processor.} Circuit cutting allows the division of a large quantum circuit into smaller, non-overlapping subcircuits, such that each can be implemented on devices with fewer qubits \cite{carrera2024combining}. \textcolor{black}{We consider a set of available quantum processors in the market denoted by $\mathcal{K} = \{1,2,\dots,K\}$. At the beginning of our DVQC framework, each processor submits a bid to take part in subcircuit execution based on its available qubit capacity and corresponding reported energy cost. Due to the incentive-based selection mechanism introduced later, not all processors are necessarily selected. Instead, the aggregator determines an optimal subset of processors $\mathcal{S} \subseteq \mathcal{K}$, where $\mathcal{S} = \{k \mid x_{k} = 1\}$ denotes the set of winning processors. Here, $x_{k}$ is the binary decision variable that represents whether processor $k$'s bid wins or not. Only the selected processors in $\mathcal{S}$ are assigned subcircuits for DVQC, while the remaining processors may participate in future bidding rounds.} A central aggregator extracts features from data, distributes the relevant features to processors according to their assigned qubits, and coordinates optimization through classical feedback. This approach is particularly beneficial \textit{when the aggregator lacks quantum processor capable of executing the full PQC} requiring a large number of qubits \cite{carrera2024combining, caleffi2024distributed}. Instead, the aggregator allocates subcircuit tasks across multiple processors, enabling parallel execution \cite{tang2021cutqc}, improved hardware utilization, and overcoming qubit count and connectivity limitations of individual devices. \textcolor{black}{We note that qubit constraints in practical quantum systems are not solely determined by the size of the large-scale PQCs. They are also influenced by hardware non-idealities such as noise, decoherence, dissipation, limited connectivity, and imperfect gate fidelity. Consequently, even small or moderately sized PQCs may face qubit limitations due to reduced reliability and usable qubit capacity in near-term quantum devices.} The steps of our proposed DVQC framework are outlined below:
\begin{enumerate}
    \item In our proposed DVQC framework, a PQC is decomposed into multiple smaller subcircuits using circuit cutting, where each subcircuit is executed independently on a dedicated quantum processor \cite{carrera2024combining}. The aggregator holds the classical input data (e.g., images, signal values, or sensor measurements), extracts feature representations, and distributes the relevant features to processors based on their assigned qubits for subcircuit execution.
    \item Each processor initializes its assigned qubits and encodes the received classical features using parameterized rotation gates before executing its designated subcircuit. The processor then applies its trainable subcircuit to the encoded quantum state. After execution, each processor performs measurements and sends the classical outcomes to the aggregator, which concatenates them into a single measurement vector.
    \item The aggregator applies classical post-processing (e.g., a linear layer followed by softmax) to generate predictions and compute the loss, which is then sent back to the processors. The processors use the loss to compute gradients and update their subcircuit parameters via stochastic gradient descent (SGD). These steps repeat until convergence. A block diagram of the proposed DVQC framework is shown in Fig.~\ref{fig:overviewblockdiagram}.
\end{enumerate}

In our framework, let each classical input data point be denoted by the feature vector
$
    \mathbf{f} \in \mathbb{R}^{d}.
$
\textcolor{black}{Here, the classical input data point corresponds to real samples drawn from a labeled training dataset such as MNIST or CIFAR-10. \textcolor{black}{Each $\mathbf{f}$ represents the feature vector of a single data sample} (for example, one image) used for supervised learning.} Each processor $k$ receives a subset of the input features denoted by \textcolor{black}{$\mathbf{f}_{k} \in \mathbb{R}^{q_{k}}$}, corresponding to its assigned qubits, where $q_{k}$ is the number of qubits handled by processor $k$. \textcolor{black}{This feature partitioning is necessary because the original full PQC would encode the entire feature vector across all qubits of a single large quantum device. Due to limited qubit availability, the aggregator distributes the feature vector across multiple selected processors according to their qubit capacities, such that each processor handles only the portion of features mapped to its assigned qubits. Consequently, every data sample is collectively processed by all selected processors.} \textcolor{black}{In this work, we adopt feature-based angle encoding, where normalized classical features directly parameterize single-qubit rotation gates. We note that in this work, we use image datasets such as MNIST and CIFAR-10. However, the encoding scheme is dataset-agnostic and can be applied to other data types such as sensor or signal measurements after appropriate feature extraction and processing.} After receiving a subset of the input features denoted by \textcolor{black}{$\mathbf{f}_{k}$}, processor $k$ initializes its qubits in the $\vert0\rangle^{\otimes q_{k}}$ state and performs classical-to-quantum encoding using an encoding unitary $E_{k}(\mathbf{f}_{k})$. The resulting quantum state on processor $k$ is given by
\begin{align}
    \vert\psi_{k}(\mathbf{f}_{k})\rangle = E_{k}(\mathbf{f}_{k}) \vert0\rangle^{\otimes q_{k}},
\end{align}
where $E_{k}(\mathbf{f}_{k})$ typically consists of parameterized single-qubit rotation gates such as $R_{x}$, $R_{y}$. \textcolor{black}{We emphasize that the encoding is performed locally at each processor using these feature-dependent rotation gates, where the classical features directly act as rotation parameters to embed task-relevant information into quantum states.} \textcolor{black}{The full PQC $U(\theta)$ is partitioned into $|\mathcal{S}|$ smaller subcircuits using circuit cutting techniques, where each subcircuit $U_{k}(\theta_{k})$ is executed on independent processor $k \in \mathcal{S}$ over disjoint qubit subsets. Since the selected processors operate on independent Hilbert spaces, the overall distributed quantum operation can be represented as the tensor product of subcircuit unitaries,
\begin{align}
    U(\theta) = \bigotimes_{k \in \mathcal{S}} U_{k}(\theta_{k}),
\end{align}
where $\theta = \{\theta_{k}\}_{k \in \mathcal{S}}$ denotes the collection of trainable parameters across all distributed subcircuits.} After classical-to-quantum encoding, the encoded quantum state at processor $k$ is given by $\vert\psi_{k}(\mathbf{f}_{k})\rangle$, which serves as the input to its assigned parameterized subcircuit. The quantum evolution at processor $k$ is therefore expressed as
\color{black}
\begin{align}
    \vert \mu_{k} \rangle = U_{k}(\theta_{k}) \vert\psi_{k}(\mathbf{f}_{k})\rangle,
\end{align}
\color{black}
where $U_{k}(\theta_{k})$ is the trainable subcircuit executed locally on processor $k \in \mathcal{S}$ with parameter vector $\theta_{k}$. 
Following quantum evolution, each processor $k \in \mathcal{S}$ performs measurements on its local qubits to generate classical outcomes. \textcolor{black}{Let $\hat{O}_{k}$ be the observable measured by processor $k$, which typically corresponds to Pauli-Z operators. For each input sample at iteration $t$, processor $k$ executes its assigned subcircuit $\phi_{k}^{(t)}$ times, (i.e., number of measurement repetitions or shots), and measurements are performed after each circuit execution. Denoting the measurement outcome of the $r$-th repetition by $o_{k}^{r}$, the measurement output is computed as the expectation value averaged over all repetitions as 
\color{black}
\begin{align}
    \mathbf{c}_{k}^{(t)} = \frac{1}{\phi_{k}^{(t)}} \sum_{r=1}^{\phi_{k}^{(t)}} o_{k}^{r} \approx \langle \mu_{k} \vert \hat{O}_{k} \vert \mu_{k} \rangle
\end{align}
\color{black}
where $\mathbf{c}_{k}^{(t)} \in \mathbb{R}^{q_{k}}$ represents the measurement vector obtained from the $q_{k}$ qubits of processor $k \in \mathcal{S}$. The averaging operation reduces the variance caused by finite sampling, i.e., the effect of quantum shot noise.} The classical measurement vectors from all selected processors are then transmitted back to the aggregator through classical communication links. \textcolor{black}{The aggregator then concatenates the received measurement vectors to form a global measurement representation given by
\begin{align}
    \mathbf{c}^{(t)} = \left[ {\mathbf{c}_{1}^{(t)}}^\top {\mathbf{c}_{2}^{(t)}}^\top \ldots {\mathbf{c}_{|\mathcal{S}|}^{(t)}}^\top \right]^\top.
\end{align}}
\textcolor{black}{Since all subcircuits correspond to disjoint partitions of the original PQC, the concatenated measurement outcomes collectively represent the same input sample.} Subsequently, the aggregator performs classical post-processing using a parameterized function $f_{\text{agg}}(\cdot;\gamma)$, such as a fully connected layer followed by a softmax activation, to produce the final prediction
$
    \hat{y} = f_{\text{agg}}(\mathbf{m};\gamma),
$
where $\gamma$ denotes the trainable parameters of the classical layer. \textcolor{black}{This post-processing stage maps the concatenated quantum measurements to task-specific outputs such as class probabilities in classification tasks, thereby linking the distributed quantum computation with a well-defined learning objective. Using the ground-truth label $y$ from the labeled dataset, the aggregator computes a training loss $\mathcal{L}(\hat{y}, y)$, where $\mathcal{L}$ can be any suitable loss function such as mean squared error or cross-entropy. Therefore, the optimization objective of the DVQC framework is to minimize the empirical training loss over the dataset.} The computed loss value is then transmitted back to the selected quantum processors. Upon receiving the loss information, each processor $k \in \mathcal{S}$ updates the parameters of its assigned subcircuit $\theta_{k}$ using gradient-based optimization such as SGD, while the aggregator simultaneously updates the classical aggregation parameters $\gamma$. We note that, for the quantum parameters, gradient estimation is performed using quantum-compatible techniques such as the parameter-shift rule. Let $t \in \{1, 2, \dots, T\}$ denote the iteration index of the DVQC training process, where $T$ is the total number of training iterations. At iteration $t$, the quantum parameters of subcircuit $k \in \mathcal{S}$ are updated using SGD as
\begin{align}
    \theta_{k}^{t+1} = \theta_{k}^{t} - \eta \nabla_{\theta_{k}} \mathcal{L}(\hat{y}, y),
\end{align}
where $\eta$ denotes the learning rate of the SGD optimizer. The above steps are executed iteratively across multiple training iterations until the model parameters converge, yielding stable predictions and minimized loss.

\subsection{DVQC Processor Energy Model}
\noindent
\textcolor{black}{In the proposed DVQC framework, each candidate quantum processor $k \in \mathcal{K}$ first estimates the total energy required to execute the assigned training workload announced by the aggregator, taking into account its available qubit resources and computational capabilities. Based on this announced workload, all processors in the market locally solve an energy minimization problem to obtain their minimum expected energy consumption for both quantum computation and measurement transmission. The minimized energy value is then reported to the aggregator as a bid, reflecting the processor’s cost of participation and its willingness to contribute computational resources in exchange for monetary incentives.}

\textit{1) Quantum Computing Energy:}
If processor $k$ is selected by the aggregator (i.e., $k \in \mathcal{S}$), it executes its assigned subcircuit on all encoded quantum states corresponding to its allocated share of training inputs at each iteration $t$. Inspired by \cite{fellous2023optimizing}, the quantum computing energy consumption of a selected processor $k \in \mathcal{S}$ over $T$ DVQC training iterations is expressed as
\begin{equation}
    E_{k}^{\text{quant}} = D_{k} q_{k} \sum_{t=1}^{T} \phi_{k}^{(t)} \left( p_{k,1}^{(t)} N_{k,1} + p_{k,2}^{(t)} N_{k,2} + p_{k,M}^{(t)} N_{k,M} \right), \label{eq7}
\end{equation}
where $D_{k}$ is the number of encoded training samples processed by processor $k$ in each DVQC iteration, and $q_{k}$ is the number of qubits involved in the subcircuit. The quantities $p_{k,1}^{(t)}$, $p_{k,2}^{(t)}$, and $p_{k,M}^{(t)}$ represent the power consumption at iteration $t$ per physical 1-qubit gate, 2-qubit gate, and measurement gate, respectively. In our model, we consider 1-qubit and 2-qubit gate operations to align with the practical limitations of near-term quantum processors \cite{tang2021cutqc}. Additionally, $N_{k,1}$, $N_{k,2}$, and $N_{k,M}$ are the average counts of these gates executed in parallel at each time step within the PQC, while $\phi_{k}^{(t)}$ corresponds to the number of noise measurement repetitions during iteration $t$ aimed at error mitigation. \textcolor{black}{In this work, we assume a fixed batch size throughout the DVQC training process, such that each processor $k$ processes a constant number of training samples $D_{k}$ at every iteration. Therefore, $D_{k}$ is independent of the iteration index $t$ and is placed outside the summation in \eqref{eq7}. For practical scenarios involving varying batch sizes, the formulation can be readily extended by replacing $D_{k}$ with the iteration-dependent quantity $D_{k}^{(t)}$ inside the summation.}

\noindent
\textit{2) Measurement Transmission Energy:}
After executing the allocated subcircuits, each quantum processor $k \in \mathcal{S}$ transmits its measurement outcomes back to the aggregator for further processing. We consider a spatial communication setup where the aggregator and the processors are located at fixed positions. The processors simultaneously upload their measurement data over a shared uplink bandwidth $B$ using power-domain non-orthogonal multiple access (NOMA). \textcolor{black}{We note that our model adopts a NOMA-based uplink communication architecture for measurement transmission, while the DVQC learning framework itself remains general and can be applied to different wireless network architectures. However, we adopt a NOMA-based uplink in this work because NOMA generally outperforms orthogonal multiple access schemes (e.g., TDMA) in terms of transmission efficiency and energy consumption.}

The wireless channel between processor $k$ and the aggregator at iteration $t$ is characterized by channel gain $h_{k}^{(t)}$, which captures large-scale path loss and small-scale Rayleigh fading. \textcolor{black}{We note that the iteration-dependent channel gain does not necessarily imply independent fading realizations across consecutive training iterations. Rather, temporal variations in small-scale fading depend on the channel coherence time and underlying propagation environment. Accordingly, the channel may remain approximately constant over multiple training iterations under slowly varying conditions, while channel variations can be incorporated when the channel conditions change.} Let us denote the noise power spectral density as $n_{0}$, then the additive white Gaussian noise power over the bandwidth $B$ is given by $\sigma^{2} = B n_{0}$. At the aggregator, SIC is employed to decode the superposed uplink signals. The received signals from processors are ordered according to their received signal strength $|h_{k}^{(t)}|^{2} p_{k,\text{tra}}^{(t)}$, where $p_{k,\text{tra}}^{(t)}$ is the uplink transmit power of processor $k$. Stronger signals are decoded first and subtracted successively during the SIC process. Under this NOMA transmission model, the achievable uplink rate of processor $k$ at iteration $t$ is expressed as
\begin{equation}
    R_{k}^{(t)}=B \log_{2}\left(1+\gamma_{k}^{(t)}\right), \label{eqn:302}
\end{equation}
where $\gamma_{k}^{(t)}$ denotes the received signal-to-interference-plus-noise ratio (SINR) of processor $k$, defined as
\begin{align}
    \gamma_{k}^{(t)} = \frac{A_{k}^{(t)}}{B_{k}^{(t)} + \sigma^{2}}. \label{eqn:306}
\end{align}
Here,
\begin{align}
    A_{k}^{(t)} = |h_{k}^{(t)}|^{2} p_{k,\text{tra}}^{(t)} \label{eqn:307}
\end{align}
represents the received signal power from processor $k$, and
\begin{align}
    B_{k}^{(t)} = \sum_{\substack{j \in \mathcal{S}, j \neq k \\ |h_{j}^{(t)}|^{2} p_{j,\text{tra}}^{(t)} < |h_{k}^{(t)}|^{2} p_{k,\text{tra}}^{(t)}}} |h_{j}^{(t)}|^{2} p_{j,\text{tra}}^{(t)} 
\end{align}
denotes the interference from weaker processors whose signals have not yet been canceled by SIC. Consequently, the transmission energy consumption for uploading the measurement outcomes of processor $k \in \mathcal{S}$ during iteration $t$ is given by
\begin{equation}
    E_{k}^{\text{trans},(t)}=\frac{s_{k}^{(t)}}{R_{k}^{(t)}}p_{k,\text{tra}}^{(t)}, \label{eqn:303}
\end{equation}
where $s_{k}^{(t)}$ denotes the size of measurement data transmitted by processor $k$. Hence, the total transmission energy of processor $k$ over $T$ iterations is calculated as
\begin{equation}
    E_{k}^{\text{trans}} = \sum_{t=1}^{T} E_{k}^{\text{trans},(t)}. \label{eqn:301}
\end{equation}
\color{black}
Therefore, the combined energy consumption of processor $k$ for quantum computing and measurement transmission over $T$ iterations is quantified as
\begin{equation}
    E_{k} = E_{k}^{\text{quant}} + E_{k}^{\text{trans}}. \label{eqn:300}
\end{equation}

At the beginning of our proposed DVQC framework, processor $k$ solves the following optimization problem to minimize its overall energy consumption for quantum computing and measurement transmission. Hence, we formulate an energy minimization problem that jointly optimizes subcircuit parameters \big($p_{k,1}^{(t)}$, $p_{k,2}^{(t)}$, $p_{k,M}^{(t)}$\big), number of noise measurements $\phi_{k}^{(t)}$, and resource allocation, i.e., transmit power \big($p_{k,\text{tra}}^{(t)}$\big).
 
\textbf{\textit{Problem 1}:}
\begin{subequations} 
\begin{align}
\min_{\begin{aligned}
    & \text{\small $p_{k,1}^{(t)}$, $p_{k,2}^{(t)}$, $p_{k,M}^{(t)}$,} \\[-0.5ex]
    & \text{\small $\phi_{k}^{(t)}$, $p_{k,\text{tra}}^{(t)}$}
\end{aligned}} \quad & E_{k} \label{eqn:17a}\\ 
\textrm{s.t.} \quad & p_{k,1}^{(t)} > 0, \quad \forall{k,t} \label{eqn:17b}\\
& p_{k,2}^{(t)} > 0, \quad \forall{k,t} \label{eqn:17c}\\
& p_{k,M}^{(t)} > 0, \quad \forall{k,t} \label{eqn:17d}\\
& p_{k,\text{tra}}^{(t)} > 0,  \quad \forall{k,t} \label{eqn:17e}\\
& \phi_{k}^{(t)} > 0, \quad \forall{k,t} \label{eqn:17f}\\
p_{k,1}^{(t)} + p_{k,2}^{(t)} + p_{k,M}^{(t)} &+ p_{k,\text{tra}}^{(t)}  \le P_{k,\text{max}}^{(t)}, \quad \forall{k,t},  \label{eqn:17g}\\
 \nonumber
\end{align} 
\end{subequations} 
\textcolor{black}{where the constraint in \eqref{eqn:17g} is imposed to ensure that, during each SCA optimization iteration, the values of the left hand side power variables remain within a feasible range and do not exceed the processor's maximum power budget, thereby maintaining physically realistic and hardware-compliant operation.}

\subsection{Incentive Model for DVQC}
\textcolor{black}{In practical DVQC deployments, quantum processors are typically owned and operated by independent entities such as academic institutions, cloud service providers, or private companies. These processors act in their self-interest and are generally unwilling to participate in distributed quantum computing tasks unless they are adequately compensated for the energy and computational resources they expend. This is particularly important for near-term quantum hardware, which is highly energy constrained. Moreover, repeated circuit execution, measurement operations, and transmission of measurement outcomes to the aggregator can incur significant energy consumption. To address this challenge, we propose a two-stage incentive mechanism for DVQC. At the beginning of the DVQC framework, the aggregator announces the quantum computing task along with its required workload to the available processors in the market. In \textit{Stage 1}, each processor locally minimizes its estimated DVQC energy cost based on the announced workload and its available qubits by solving the processor-side energy minimization problem (i.e., \textbf{\textit{Problem 1}}) that jointly optimizes subcircuit execution parameters, the number of noise measurements, and communication transmit power. The minimized energy cost is then reported to the aggregator as the processor's bid. In \textit{Stage 2}, the aggregator uses the reported energy costs to solve a system utility maximization problem to select an optimal subset of processors under payment budget and qubit coverage constraints.}

\textcolor{black}{In practical quantum computing market, multiple processors compete to execute portions of the PQC in order to earn these incentives \cite{carrera2024combining}, creating a competitive environment that encourages efficient use of computational resources and reliable task execution. In our framework, the aggregator offers processor $k$ a monetary payment $m_{k}$ as a compensation for executing its assigned subcircuit and transmitting measurement outcomes for $T$ iterations. Each processor $k$ reports its estimated minimized total energy cost $E_{k}$ to the aggregator. Based on these reported energy costs and available qubit capacities of the processors, the aggregator determines the set of winner processors by solving a system utility maximization problem. The objective is to maximize the net system utility, defined as the aggregate difference between the offered payments and the energy costs of the processors, while satisfying the payment budget and required qubit coverage constraints. Specifically, at the beginning of the DVQC framework, the aggregator solves the following optimization problem:}

\textbf{\textit{Problem 2}:}
\begin{subequations} 
\begin{align}
\max_{x_{k}} \quad & \sum_{k=1}^{K} \bigg(m_{k} - E_{k}\bigg) x_{k} \label{eqn:18a}\\ 
\textrm{s.t.} \quad & x_{k} \in \{0, 1\}, \quad \forall{k}, \label{eqn:18b}\\
& \sum_{k=1}^{K} m_{k} x_{k} \le M_{\text{max}}, \quad \forall{k}, \label{eqn:18c}\\
& \sum_{k=1}^{K} q_{k} x_{k} \ge Q, \quad \forall{k}, \label{eqn:18d}
\end{align} 
\end{subequations} 
where $x_{k}$ is the binary decision variable indicating whether processor $k$ is selected by the aggregator to participate in subcircuit execution (i.e., $x_{k} =1$ if selceted, and $x_{k} = 0$ otherwise), and \eqref{eqn:18b} is the corresponding binary constraint. Moreover, $M_{\text{max}}$ in \eqref{eqn:18c} is the total payment budget the aggregator can allocate. Furthermore, to guarantee sufficient qubit coverage in the DVQC framework, we impose the constraint \eqref{eqn:18d}, where $q_{k}$ is the number of qubits in processor $k$ and $Q$ is the total number of qubits in the original PQC. This ensures the selected subcircuits collectively cover the entire PQC. From \textbf{\textit{Problem 2}}, it is clear that the aggregator does not pre-define how qubits are divided among subcircuits; instead, this partitioning emerges as a result of the system utility maximization process. 

\textcolor{black}{We note that, our proposed incentive mechanism is closely tied to the energy optimization performed by each processor. To increase their chances of selection and maximize system utility, processors minimize their reported energy costs by solving the processor-side energy minimization problem (i.e., Problem 1). Using these reported energy values, the aggregator then solves the system utility maximization problem (i.e., Problem 2) to select an optimal subset of processors for DVQC execution. Therefore, Problem 1 and Problem 2 are interdependent in our framework. Since the aggregator selects processors based on the difference between offered payment and reported energy cost, processors that operate in a more energy-efficient way are more likely to be selected and compensated. This interaction establishes a natural equilibrium between processor incentives and aggregator objectives, where processors minimize energy to improve their selection probability while the aggregator maximizes system utility under budget and qubit constraints. Consequently, the proposed two-stage framework achieves a balanced trade-off between processor compensation and system efficiency.}

\section{Proposed Solution} \label{sectionIII}
\subsection{Energy Optimization using BCD and SCA}
\textit{\textbf{Problem 1}} is difficult to solve directly because of its non-convex objective function, making it unsuitable for standard convex optimization techniques. To overcome this, we use block coordinate descent (BCD) and successive convex approximation (SCA) techniques that splits the problem into two smaller, tractable subproblems. In particular, we divide the control variables into two groups: ($p_{k,1}^{(t)}$, $p_{k,2}^{(t)}$, $p_{k,M}^{(t)}$, $\phi_{k}^{(t)}$) and ($p_{k,\text{tra}}^{(t)}$). The two subproblems are solved in an iterative manner until the convergence is achieved.

\noindent
\textbf{\textit{Sub-problem 1 (Joint processor subcircuit power control and number of noise measurements optimization)}:}
\begin{subequations} 
\begin{align} 
\min_{\begin{aligned}
    & \text{\small $p_{k,1}^{(t)}$, $p_{k,2}^{(t)}$, $p_{k,M}^{(t)}$, $\phi_{k}^{(t)}$} \\[-0.5ex]
\end{aligned}} \quad & E_{k} \label{eqn:19a'}\\ 
\textrm{s.t.} \quad & p_{k,1}^{(t)} > 0, \quad \forall{k,t} \label{eqn:19b'}\\
& p_{k,2}^{(t)} > 0, \quad \forall{k,t} \label{eqn:19c'}\\
& p_{k,M}^{(t)} > 0, \quad \forall{k,t} \label{eqn:19d'}\\
& \phi_{k}^{(t)} > 0, \quad \forall{k,t} \label{eqn:19e'}\\
p_{k,1}^{(t)} + p_{k,2}^{(t)} + p_{k,M}^{(t)} &+ p_{k,\text{tra}}^{(t)}  \le P_{k,\text{max}}^{(t)}, \quad \forall{k,t}.  \label{eqn:19f'}\\
 \nonumber
\end{align} \label{eqn:19'}
\end{subequations}

\noindent
\textbf{\textit{Sub-problem 2 (Processor transmit power optimization)}:}
\begin{subequations} 
\begin{align}
\min_{\begin{aligned}
    & \text{\small $p_{k,\text{tra}}^{(t)}$}
\end{aligned}} \quad & E_{k} \label{eqn:20a}\\ 
\textrm{s.t.} \quad & p_{k,\text{tra}}^{(t)} > 0,  \quad \forall{k,t} \label{eqn:20b}\\
p_{k,1}^{(t)} + p_{k,2}^{(t)} + p_{k,M}^{(t)} &+ p_{k,\text{tra}}^{(t)}  \le P_{k,\text{max}}^{(t)}, \quad \forall{k,t}.  \label{eqn:20c}\\
 \nonumber
\end{align} \label{eqn:20}
\end{subequations}

\noindent
\(\bigstar\) \textit{Sub-problem 1 (Joint processor's subcircuit power control and number of noise measurements optimization)}:

\textit{\textbf{Sub-problem 1}} in \eqref{eqn:19'} is inherently non-convex because of the objective function in \eqref{eqn:19a'}. Consequently, our next step is to transform \eqref{eqn:19'} into a convex formulation.

\noindent
\underline{\textit{\textbf{\eqref{eqn:19a'}}}:} We express the objective function as 
\begin{align}
    E_{k} = &D_{k} q_{k} \sum_{t=1}^{T} \phi_{k}^{(t)} \left( p_{k,1}^{(t)} N_{k,1} + p_{k,2}^{(t)} N_{k,2} + p_{k,M}^{(t)} N_{k,M} \right) \nonumber \\
    &\hspace{10em} + E_{k}^{\text{trans}}. \label{eqn:21}
\end{align}
The first, second, and third terms in the objective function of \eqref{eqn:21} is non-convex; therefore, we proceed to apply convexification techniques to these terms. \textit{For the first term of the objective function,} and $\phi_{k}^{(t)} > 0$ and $p_{k,1}^{(t)} > 0$, we leverage SCA to approximate $\phi_{k}^{(t)} p_{k,1}^{(t)}$ as
\begin{align}
    \phi_{k}^{(t)} p_{k,1}^{(t)} \leq \frac{1}{2} \frac{{p_{k,1}^{(t)}}_{i}}{{\phi_{k}^{(t)}}_{i}} {\phi_{k}^{(t)}}^2 + \frac{1}{2} \frac{{\phi_{k}^{(t)}}_{i}}{{p_{k,1}^{(t)}}_{i}} {p_{k,1}^{(t)}}^2,
\end{align}
where $i$ denotes the SCA iteration index. Similarly, for $p_{k,2}^{(t)} > 0$ and $p_{k,3}^{(t)} > 0$, we convexity the second and third terms as
$
    \phi_{k}^{(t)} p_{k,2}^{(t)} \leq \frac{1}{2} \frac{{p_{k,2}^{(t)}}_{i}}{{\phi_{k}^{(t)}}_{i}} {\phi_{k}^{(t)}}^2 + \frac{1}{2} \frac{{\phi_{k}^{(t)}}_{i}}{{p_{k,2}^{(t)}}_{i}} {p_{k,2}^{(t)}}^2,
$
$
    \phi_{k}^{(t)} p_{k,3}^{(t)} \leq \frac{1}{2} \frac{{p_{k,3}^{(t)}}_{i}}{{\phi_{k}^{(t)}}_{i}} {\phi_{k}^{(t)}}^2 + \frac{1}{2} \frac{{\phi_{k}^{(t)}}_{i}}{{p_{k,3}^{(t)}}_{i}} {p_{k,3}^{(t)}}^2.
$

\noindent
After convexification, \textit{\textbf{Sub-problem 1}} is written as follows.
\noindent
\textbf{\textit{Sub-problem 1 (convex equivalent)}:}
\begin{subequations} 
\begin{align} 
\min_{\begin{aligned} 
    & \text{\small $p_{k,1}^{(t)}$, $p_{k,2}^{(t)}$,} \\[-0.5ex]
    & \text{\small $p_{k,M}^{(t)}$, $\phi_{k}^{(t)}$}
\end{aligned}} &\quad D_{k} q_{k} \sum_{t=1}^{T} \bigg[ \bigg(\frac{1}{2} \frac{{p_{k,2}^{(t)}}_{i}}{{\phi_{k}^{(t)}}_{i}} {\phi_{k}^{(t)}}^2 + \frac{1}{2} \frac{{\phi_{k}^{(t)}}_{i}}{{p_{k,2}^{(t)}}_{i}} {p_{k,2}^{(t)}}^2\bigg) N_{k,1} \nonumber \\
&\hspace{2em}+ \bigg(\frac{1}{2} \frac{{p_{k,3}^{(t)}}_{i}}{{\phi_{k}^{(t)}}_{i}} {\phi_{k}^{(t)}}^2 + \frac{1}{2} \frac{{\phi_{k}^{(t)}}_{i}}{{p_{k,3}^{(t)}}_{i}} {p_{k,3}^{(t)}}^2\bigg) N_{k,2} \nonumber \\
&\hspace{-2em}+ \bigg(\frac{1}{2} \frac{{p_{k,M}^{(t)}}_{i}}{{\phi_{k}^{(t)}}_{i}} {\phi_{k}^{(t)}}^2 + \frac{1}{2} \frac{{\phi_{k}^{(t)}}_{i}}{{p_{k,M}^{(t)}}_{i}} {p_{k,M}^{(t)}}^2\bigg) N_{k,M} \bigg] + E_{k}^{\text{trans},(t)} \label{eqn:23a'}\\ 
\textrm{s.t.} \quad & \eqref{eqn:19b'}, \eqref{eqn:19c'}, \eqref{eqn:19d'}, \eqref{eqn:19e'}, \eqref{eqn:19f'}. \label{eqn:23b'}
\end{align} \label{eqn:23'}
\end{subequations}
The \textit{\textbf{problem in \eqref{eqn:23'}}} is a standard convex optimization problem that can be solved efficiently using tools such as CVX.

\noindent
\textbf{\textit{Complexity Analysis of Sub-problem 1}:} This problem contains $4KT$ scalar decision variables and $5KT$ linear constraints, resulting in a per-iteration computational complexity approximately proportional to $\mathcal{O}\big((4KT)^2 \sqrt{5KT}\big)$ \cite{ben2001lectures}.

\noindent
\(\bigstar\) \textit{Sub-problem 2 (Processor's transmit power optimization)}:

\textit{\textbf{Sub-problem 2}} in \eqref{eqn:20} in non-convex because of the objective function in \eqref{eqn:20a}. Hence, we now convexify \eqref{eqn:20a}.

\noindent
\underline{\textit{\textbf{\eqref{eqn:20a}}}:} \textcolor{black}{From \eqref{eqn:300}, we have
\begin{equation}
    E_{k} = E_{k}^{\text{quant}} + E_{k}^{\text{trans}}. \label{eqn:304}
\end{equation}
Substituting \eqref{eqn:301} and \eqref{eqn:303} into \eqref{eqn:304}, we get
\begin{equation}
    E_{k} = E_{k}^{\text{quant}} + \sum_{t=1}^{T} \frac{s_{k}^{(t)}}{R_{k}^{(t)}}p_{k,\text{tra}}^{(t)}. \label{eqn:305}
\end{equation}
Substituting \eqref{eqn:302}, \eqref{eqn:306}, and \eqref{eqn:307} into \eqref{eqn:305}, we write 
\begin{align}
    E_{k} = &E_{k}^{\text{quant}} + \sum_{t=1}^{T} \frac{s_{k}^{(t)}}{B \log_{2}\left(1+\frac{|h_{k}^{(t)}|^{2} p_{k,\text{tra}}^{(t)}}{B_{k}^{(t)} + \sigma^{2}}\right)}p_{k,\text{tra}}^{(t)}. \label{eqn:21'}
\end{align}}
The second term in \eqref{eqn:21'} is non-convex because of the presence of the $\log$ function. \textit{For the second term of the objective function,} we introduce a slack variable $\alpha$ defined as
\begin{align}
    \frac{s_{k}^{(t)}}{B \log_{2}\left(1+\frac{|h_{k}^{(t)}|^{2} p_{k,\text{tra}}^{(t)}}{B_{k}^{(t)} + \sigma^{2}}\right)} \leq \alpha. \label{eqn:25'}
\end{align}
Next, we introduce an additional slack variable $\beta$, and reformulate \eqref{eqn:25'} as follows:
\begin{subnumcases}{\eqref{eqn:25'}\Leftrightarrow}
   s_{k}^{(t)} \leq \alpha \beta, \label{eqn:26a'} \\
   B \log_{2}\left(1+\frac{|h_{k}^{(t)}|^{2} p_{k,\text{tra}}^{(t)}}{B_{k}^{(t)} + \sigma^{2}}\right) \geq \beta. \label{eqn:26b'}
\end{subnumcases}
Equation \eqref{eqn:25'} is equivalent to this system of inequalities. \textcolor{black}{This reformulation is adopted to decouple the original non-convex fractional term that involves the logarithmic rate expression under the NOMA-based uplink model. As we aim to minimize the objective function, i.e., the energy consumption, we introduce $\alpha$ as an upper bound on the transmission time component $(s_{k}^{(t)}/B \log_{2}(1+\frac{|h_{k}^{(t)}|^{2} p_{k,\text{tra}}^{(t)}}{B_{k}^{(t)} + \sigma^{2}}))$, while $\beta$ serves as a lower bound on the achievable transmission rate $\log_{2}(1+(|h_{k}^{(t)}|^{2} p_{k,\text{tra}}^{(t)}/(B_{k}^{(t)} + \sigma^{2})))$. By introducing $\alpha$ and $\beta$, the non-convex fraction is transformed into two constraints: a bilinear constraint \eqref{eqn:26a'}, and logarithmic constraint \eqref{eqn:26b'}. This decomposition enables us to handle the bilinear and logarithmic terms separately and apply SCA to obtain a tractable convex surrogate at each iteration, while preserving equivalence to the original non-convex objective function.}

Next, we investigate the convexity properties of the inequalities within this system.

\noindent
\underline{\textit{\textbf{\eqref{eqn:26a'}}}:} Using first-order Taylor expansion, we approximate \eqref{eqn:26a'} as
\begin{align}
    \frac{1}{4} \bigg[(\alpha_{i}+\beta_{i})^2 &+ 2 (\alpha_{i}+\beta_{i})(\alpha+\beta-\alpha_{i}-\beta_{i})\bigg] \nonumber \\
    &- s_{k}^{(t)} \geq \frac{1}{4} (\alpha-\beta)^2. \label{eqn:29'}
\end{align}
\textcolor{black}{The detailed convexification process is provided in Appendix~\ref{appendixA}.}

\noindent
\underline{\textit{\textbf{\eqref{eqn:26b'}}}:} Using the inequality
$
    \ln(1+\psi) \ge \ln(1+\psi_{i}) + \frac{\psi_{i}}{\psi_{i}+1} - \frac{(\psi_{i})^{2}}{\psi_{i}+1} \frac{1}{\psi}
$
\cite{nguyen2017precoder}, the left-hand side of \eqref{eqn:26b'} is written as follows:
\begin{align}
    \ln(1+{\delta}_{i}) + \frac{{\delta}_{i}}{{\delta}_{i}+1} - \frac{({\delta}_{i})^{2}}{{\delta}_{i}+1} \frac{1}{\delta} \geq \frac{\beta \ln{2}}{B}, \label{eqn:35'}
\end{align}
where $\delta$ and $\delta_{i}$ are defined as $\delta = \frac{|h_{k}^{(t)}|^{2} p_{k,\text{tra}}^{(t)}}{B_{k}^{(t)} + \sigma^{2}}$ and $\delta_{i} = \frac{{|h_{k}^{(t)}|^{2} p_{k,\text{tra}}^{(t)}}_{i}}{B_{k}^{(t)} + \sigma^{2}}$. For $\alpha > 0$ and $p_{k,\text{tra}}^{(t)} > 0$, we use SCA to approximate $\alpha p_{k,\text{tra}}^{(t)}$ as
\begin{align}
    \alpha p_{k,\text{tra}}^{(t)} \leq \frac{1}{2} \frac{{p_{k,\text{tra}}^{(t)}}_{i}}{{\alpha}_{i}} {\alpha}^2 + \frac{1}{2} \frac{{\alpha}_{i}}{{p_{k,\text{tra}}^{(t)}}_{i}} {p_{k,\text{tra}}^{(t)}}^2. \label{eqn:38new}
\end{align}
\noindent
After convexification, \textit{\textbf{Sub-problem 2}} is written as follows.

\noindent
\textbf{\textit{Sub-problem 2 (convex equivalent)}:}
\begin{subequations} 
\begin{align}
\min_{\begin{aligned}
    & \text{\small $p_{k,\text{tra}}^{(t)}$}
\end{aligned}} \quad & E_{k}^{\text{quant}} + \sum_{t=1}^{T} \bigg(\frac{1}{2} \frac{{p_{k,\text{tra}}^{(t)}}_{i}}{{\alpha}_{i}} {\alpha}^2 + \frac{1}{2} \frac{{\alpha}_{i}}{{p_{k,\text{tra}}^{(t)}}_{i}} {p_{k,\text{tra}}^{(t)}}^2\bigg) \label{eqn:33a'}\\ 
\textrm{s.t.} \quad & \eqref{eqn:29'}, \eqref{eqn:35'}, \eqref{eqn:20b}, \eqref{eqn:20c}. \label{eqn:33b'}
\end{align} \label{eqn:33'}
\end{subequations}
The \textit{\textbf{problem in \eqref{eqn:33'}}} is a standard convex optimization problem that can be solved efficiently using tools such as CVX. The procedure for solving \textbf{\textit{Problem 1}} using the proposed BCD-SCA method is summarized in Algorithm~\ref{SCAalg}.

\noindent
\textbf{\textit{Complexity Analysis of Sub-problem 2}:} This problem contains $(KT)$ scalar decision variables and $(4KT)$ linear constraints, resulting in a per-iteration computational complexity approximately proportional to $\mathcal{O}\big((KT)^2 \sqrt{4KT}\big)$ \cite{ben2001lectures}.

\makeatletter
\renewcommand{\Statex}{\item[]\hskip\ALG@thistlm}
\makeatother

\begin{algorithm}[t]
  \caption{SCA-based Joint Optimization Algorithm}
  \label{SCAalg}
\begin{algorithmic}[1]
 \footnotesize
    \Statex \textbf{Input:} 
            \Statex Set the iteration index $i=0$;
            \Statex Define a feasible initial solution
            ${p_{k,1}^{(t)}}_{0}$, ${p_{k,2}^{(t)}}_{0}$, ${p_{k,M}^{(t)}}_{0}$, ${\phi_{k}^{(t)}}_{0}$, ${p_{k,\text{tra}}^{(t)}}_{0}$ for Problem 1;
    \Statex \textbf{Repeat}
            \Statex Set $i \gets i+1$
            \Statex Solve \textbf{ Sub-problem 1 (Equivalent)} to obtain ${p_{k,1}^{(t)}}_{i}$, ${p_{k,2}^{(t)}}_{i}$, ${p_{k,M}^{(t)}}_{i}$, ${\phi_{k}^{(t)}}_{i}$;
            \Statex Solve \textbf{ Sub-problem 2 (Equivalent)} to obtain ${p_{k,\text{tra}}^{(t)}}_{i}$;
    \Statex \textbf{Until} convergence. 
    \Statex \textbf{Output:}
            \Statex Optimal ${p_{k,1}^{(t)}}^{*}$, ${p_{k,2}^{(t)}}^{*}$, ${p_{k,M}^{(t)}}^{*}$, ${\phi_{k}^{(t)}}^{*}$, ${p_{k,\text{tra}}^{(t)}}^{*}$.
\end{algorithmic} 
\end{algorithm}

\subsection{ADMM-based Winner Selection Optimization}
The winner selection optimization problem, i.e., \textbf{\textit{Problem 2}}, maximizes a linear objective function subject to binary selection constraints and a global payment budget. The binary (integer) selection constraints make the feasible set discrete and non-convex. Although the objective function is linear, these binary constraints cause the overall winner selection problem to become a \textit{non-convex combinatorial optimization problem} that is \textit{NP-hard} and computationally challenging to solve optimally. To efficiently solve the problem, we adopt the \textit{alternating direction method of multipliers (ADMM)} framework. 

\noindent
\textit{\textbf{Variable updates of ADMM:}} To enable tractable optimization, we first relax the binary constraints to continuous box constraints and then apply variable splitting to decompose the problem into simpler subproblems. Specifically, the binary variable $x_{k} \in \{0, 1\}$ is relaxed to continuous variable $x_{k} \in [0, 1]$, enabling a convex relaxation of the original problem. We now introduce an auxiliary variable $z_{k}$ to separate the budget constraint from the objective function, leading to the following equivalent reformulation:

\begin{subequations} 
\begin{align}
\min_{x_{k}, z_{k}} \quad & - \sum_{k=1}^{K} \bigg(m_{k} - E_{k}\bigg) x_{k} \label{eqn:19a}\\ 
\textrm{s.t.} \quad & x_{k} \in \{0, 1\}, \quad \forall{k}, \label{eqn:19b}\\
& \sum_{k=1}^{K} m_{k} z_{k} \le M_{\text{max}}, \label{eqn:19c}\\
& \sum_{k=1}^{K} q_{k} z_{k} \ge Q, \quad \forall{k}, \label{eqn:200}\\
& x_{k} = z_{k}, \quad \forall{k}. \label{eqn:19d}
\end{align} 
\end{subequations} 
Since ADMM is inherently designed to solve minimization problems, we minimize the negative of the original maximization objective function. We denote the inner iteration index within the ADMM algorithm itself as $i$. Specifically, before starting the DVQC training, ADMM runs several update iterations indexed by $i = 1, 2, 3, \dots, I$ until convergence for winner selection optimization. The augmented Lagrangian function at ADMM iteration $i$ is defined as
\begin{align}
    \mathcal{L}_{\rho} (x_{k}^{(i)}, z_{k}^{(i)}, \lambda_{k}^{(i)}) &= - \sum_{k=1}^{K} \bigg(m_{k} - E_{k}\bigg) x_{k}^{(i)} \nonumber \\
    &+ \frac{\rho}{2} \sum_{k=1}^{K} {\bigg(x_{k}^{(i)} - z_{k}^{(i)} + \lambda_{k}^{(i)}\bigg)}^{2}, 
\end{align}
where $\rho > 0$ is a penalty parameter and $\lambda_{k}^{(i)}$ is a scaled dual variable associated with the equality constraint. In ADMM, the augmented Lagrangian combines the objective function and the constraints using dual variables and penalty terms, enabling decomposition of the problem into simpler subproblems while enforcing constraint satisfaction. Now, at ADMM iteration $i+1$, the algorithm updates the variables as follows: 

\noindent
\textit{\textbf{1) Update of $x_{k}$}}: Fixing $z_{k}^{(i)}$ and $\lambda_{k}^{(i)}$, variable $x_{k}$ is updated 
\begin{align}
    x_{k}^{(i+1)} = \min \bigg(1, \max \bigg(0, z_{k}^{(i)} - \lambda_{k}^{(i)} &+ \frac{m_{k} - E_{k}}{\rho}\bigg)\bigg), \nonumber \\ 
    &\forall{k \in \mathcal{K}}.
\end{align}
\textcolor{black}{The detailed derivation of the update for $x_{k}$ is provided in Appendix~\ref{appendixB}.}

\noindent
\textit{\textbf{2) Update of $z_{k}$}}: Fixing $x_{k}^{(i)}$ and $\lambda_{k}^{(i)}$, variable $z_{k}^{(i)}$ is updated by solving the following optimization problem at iteration $i+1$:
\begin{align}
    z_{k}^{(i+1)} &= \arg \min_{z_{k}} \bigg\{\frac{\rho}{2} \| x_{k}^{(i+1)} - z_{k} + \lambda_{k}^{(i)}\|_{2}^{2}\bigg\} \nonumber \\
    \textrm{s.t.} \quad & \sum_{k=1}^{K} m_{k} z_{k} \le M_{\text{max}}, \quad \forall{t}, \nonumber \\
    & \sum_{k=1}^{K} q_{k} z_{k} \ge Q, \nonumber \\ 
    & 0 \leq z_{k} \leq 1, \quad \forall{k}. \label{eqn:25}
\end{align}
More specifically, the problem in \eqref{eqn:25} reduces to finding the Euclidean projection of $\mathbf{v}$ onto the half-space
\begin{align}
    \mathcal{C} = \bigg\{\mathbf{z} | {\mathbf{m}}^{\top} \mathbf{z} \le M_{\text{max}},  {\mathbf{q}}^{\top} \mathbf{z} \ge Q, 0 \leq \mathbf{z} \leq 1 \bigg\}.
\end{align}
We note that $\mathbf{v}$, $\mathbf{x}^{(i+1)}$, \textcolor{black}{$\boldsymbol{\lambda}^{(i)}$}, $\mathbf{m}$, $\mathbf{z}$, and $\mathbf{q}$ are vectors defined as $\mathbf{v} = [v_{1}, v_{2}, \dots, v_{K}]$, $\mathbf{x}^{(i+1)} = [x_{1}^{(i+1)}, x_{2}^{(i+1)}, \dots, x_{K}^{(i+1)}]$, \textcolor{black}{$\boldsymbol{\lambda}^{(i)} = [\lambda_{1}^{i}, \lambda_{2}^{i}, \dots, \lambda_{K}^{i}]$}, $\mathbf{m} = [m_{1}, m_{2}, \dots, m_{K}]$, $\mathbf{z} = [z_{1}, z_{2}, \dots, z_{K}]$, and \textcolor{black}{$\mathbf{q} = [q_{1}, q_{2}, \dots, q_{K}]$}, representing the vectors of auxiliary variables ($\mathbf{v}$), processor selection variables, dual variables, payment offers, auxiliary variables ($\mathbf{z}$), and qubit capacities, respectively. If $\mathbf{v} \in \mathcal{C}$, i.e.,
$
    {\mathbf{m}}^{\top} \mathbf{z} \le M_{\text{max}}, {\mathbf{q}}^{\top} \mathbf{z} \ge Q, 0 \leq \mathbf{z} \leq 1,
$
then the projection is
$
    \mathbf{z}^{(i+1)} = \mathbf{v},
$
where $\mathbf{z}^{(i+1)} = [z^{(i+1)}_{1}, z^{(i+1)}_{2}, \dots, z^{(i+1)}_{K}]$. Otherwise, $\mathbf{z}^{(i+1)}$ is obtained as the solution to the convex quadratic programming problem in \eqref{eqn:25}, which finds the point in $\mathcal{C}$ closest to $\mathbf{v}$ in Euclidean distance. This projection is generally computed using numerical solvers specialized for QP. In this work, we solve the problem in \eqref{eqn:25} using the CVXPY optimization library, which efficiently finds the closest feasible point to $\mathbf{v}$ under the budget, qubit coverage, and box constraints.

\textcolor{black}{We note that at ADMM iteration $(i+1)$, the optimization variables are updates sequentially in the order $x_{k}$, $z_{k}$, and $\lambda_{k}$. Therefore, when updating $z_{k}^{(i+1)}$, the variable $x_{k}^{(i+1)}$ has already been updated, while $\lambda_{k}^{(i+1)}$ is not yet available and remains at its previous value $\lambda_{k}^{(i)}$. Consequently, the quadratic term in \eqref{eqn:25} depends on $x_{k}^{(i+1)}$ and $\lambda_{k}^{(i)}$, which motivates defining an auxiliary variable (vector) $\mathbf{v}$ as
\begin{align}
    \mathbf{v} := \mathbf{x}^{(i+1)} + \textcolor{black}{\boldsymbol{\lambda}^{(i)}}.
\end{align}
Since the objective function in \eqref{eqn:25} contains only the quadratic penalty term $\frac{\rho}{2} \| x_{k}^{(i+1)} - z_{k} + \lambda_{k}^{(i)}\|_{2}^{2}$, the update of $\mathbf{z}$ corresponds to minimizing the squared Euclidean distance between $\mathbf{z}$ and the intermediate point $\mathbf{x}^{(i+1)} + \textcolor{black}{\boldsymbol{\lambda}^{(i)}}$. Therefore, the problem is equivalently interpreted as projecting $\mathbf{v} = \mathbf{x}^{(i+1)} + \textcolor{black}{\boldsymbol{\lambda}^{(i)}}$ onto the feasible set defined by the budget constraint, qubit capacity constraint, and box constraint. Clearly, the objective is strictly convex and the constraints are linear. Therefore, the resulting problem is a convex quadratic program (QP) with a unique global solution.}


\textit{\textbf{3) Update of $\lambda_{k}$}}: After updating $x_{k}$ and $z_{k}$ for all $k \in \mathcal{K}$, the variable $\lambda_{k}$ is updated as
\begin{align}
    \lambda_{k}^{(i+1)} = \lambda_{k}^{(i)} + x_{k}^{(i+1)} - z_{k}^{(i+1)}, \quad \forall{k \in \mathcal{K}}. \label{32}
\end{align}
From \eqref{32}, it is clear that if $x_{k}^{(i+1)}$ and $z_{k}^{(i+1)}$ are close, $\lambda_{k}^{(i)}$ changes less, indicating the constraint \eqref{eqn:19d} is being better satisfied. 

\noindent
\textit{\textbf{Convergence criteria of ADMM}}: The ADMM algorithm iterates until:
\begin{itemize}
    \item The primal residual measures how close $\mathbf{x}$ and $\mathbf{z}$ are as
    $
        \|\mathbf{x}^{i+1} - \mathbf{z}^{i+1}\|_{2}.
    $
    \item The dual residual measures how much $\mathbf{z}$ changes between iterations, scaled by penalty $\rho$ as
    $
        \rho \|\mathbf{z}^{i+1} - \mathbf{z}^{i}\|_{2}. 
    $
\end{itemize}
The algorithm terminates when both primal and dual residuals fall below predefined thresholds $\epsilon_{\text{prim}}$, $\epsilon_{\text{dual}}$ as:
\begin{align}
    \begin{cases}
        \|\mathbf{x}^{i+1} - \mathbf{z}^{i+1}\|_{2} \leq \epsilon_{\text{prim}}\\
        \rho \|\mathbf{z}^{i+1} - \mathbf{z}^{i}\|_{2} \leq \epsilon_{\text{dual}}.
    \end{cases}
\end{align}
When both residuals fall below the preset small thresholds, the algorithm is considered to have converged, indicating:
\begin{itemize}
    \item Primal feasibility: $\mathbf{x} \approx \mathbf{z}$.
    \item Dual feasibility: The updates have stabilized.
\end{itemize}

\noindent
\textit{\textbf{Final rounding step of ADMM}}: Because the binary constraint $x_{k} \in \{0, 1\}$ was relaxed to continuous constraint $x_{k} \in [0, 1]$, the solution $x_{k}$ at convergence is not binary. To obtain a feasible binary solution, a thresholding or rounding step is applied as
\begin{align}
    {x_{k}}^* = 
    \begin{cases}
    1, & \text{if } x_k^{(i)} \geq \tau, \\
    0, & \text{otherwise},
    \end{cases}
    \quad \forall k \in \mathcal{K},
\end{align}
where $\tau$ is a threshold, typically set to $0.5$. We mention that \textit{our ADMM approach differs from standard ADMM} by incorporating complex, problem-specific constraints, including total payment budget, qubit coverage, and box constraints. While standard ADMM often uses simple, closed-form updates for unconstrained or lightly constrained problems \cite{wang2025max}, our formulation requires solving a constrained quadratic program during the projection step, i.e., update of $z_{k}$. This projection ensures feasibility under all incentive and resource constraints and is handled numerically using CVXPY, making our method more specialized and tailored than generic ADMM.

\noindent
\textit{\textbf{Complexity Analysis of ADMM}}: The computational complexity of the ADMM-based incentive optimization is dominated by the $z_{k}$-update , which requires solving a convex quadratic program with complexity $\mathcal{O}(K^{3})$ per ADMM iteration, while the updates of $x_{k}$ and $\lambda_{k}$ are both of complexity $\mathcal{O}(K)$ \cite{chang2016asynchronous}. Overall, the total complexity is $\mathcal{O}(IK^{3})$, where $I$ is the number of ADMM iterations. \textcolor{black}{In practice, the ADMM-based winner selection algorithm converges within a very small number of iterations (e.g., around 3-5 iterations as observed in Fig.~\ref{fig:energy4schemes}) and is executed only once prior to DVQC training, resulting in negligible overhead compared to the overall learning process. Moreover, unlike exhaustive search with exponential complexity $\mathcal{O}(2^{K})$, the proposed method achieves scalable polynomial-time complexity $\mathcal{O}(IK^{3})$.}

\section{Simulation Results and Evaluation} \label{sectionIV}

\subsection{Parameter Setting for DVQC framework}
\textcolor{black}{We evaluate our proposed DVQC framework using TorchQuantum across two benchmark datasets: MNIST \cite{lecun1998gradient} and CIFAR-10 \cite{krizhevsky2009learning}.} To simulate circuit cutting, we use two partitioning configurations of PQCs. In the first configuration, a 4-qubit PQC is decomposed into two 2-qubit subcircuits, each executed independently by a quantum processor. In the second configuration, a 6-qubit PQC is partitioned into three subcircuits consisting of 1, 2, and 3 qubits, respectively. For both configurations, classical input features are encoded into quantum states via single-qubit $R_{x}$ gates. The encoded quantum state is then distributed across multiple quantum processors, each executing its designated subcircuit. All quantum subcircuits contain trainable single-qubit rotation gates, i.e., $R_{x}$ and $R_{y}$, followed by Pauli-Z measurements. Measurement results from all subcircuits are aggregated and processed by a classical linear layer, followed by a log-softmax activation function for final classification. All models are trained using the SGD optimizer with a learning rate of 0.001. We train the models for 50 epochs using batch sizes of 64 for training. \textcolor{black}{The models are evaluated on previously unseen test samples using a batch size of 64 to assess their classification accuracy and loss, where $80\%$ of the MNIST and CIFAR-10 datasets are used for training and the remaining $20\%$ for testing.} Experiments are conducted on GPU-enabled environments.

\begin{figure}[t!]
    \centering
    \footnotesize
    \begin{subfigure}[t]{0.49\linewidth} 
        \centering
        \includegraphics[width=\linewidth]{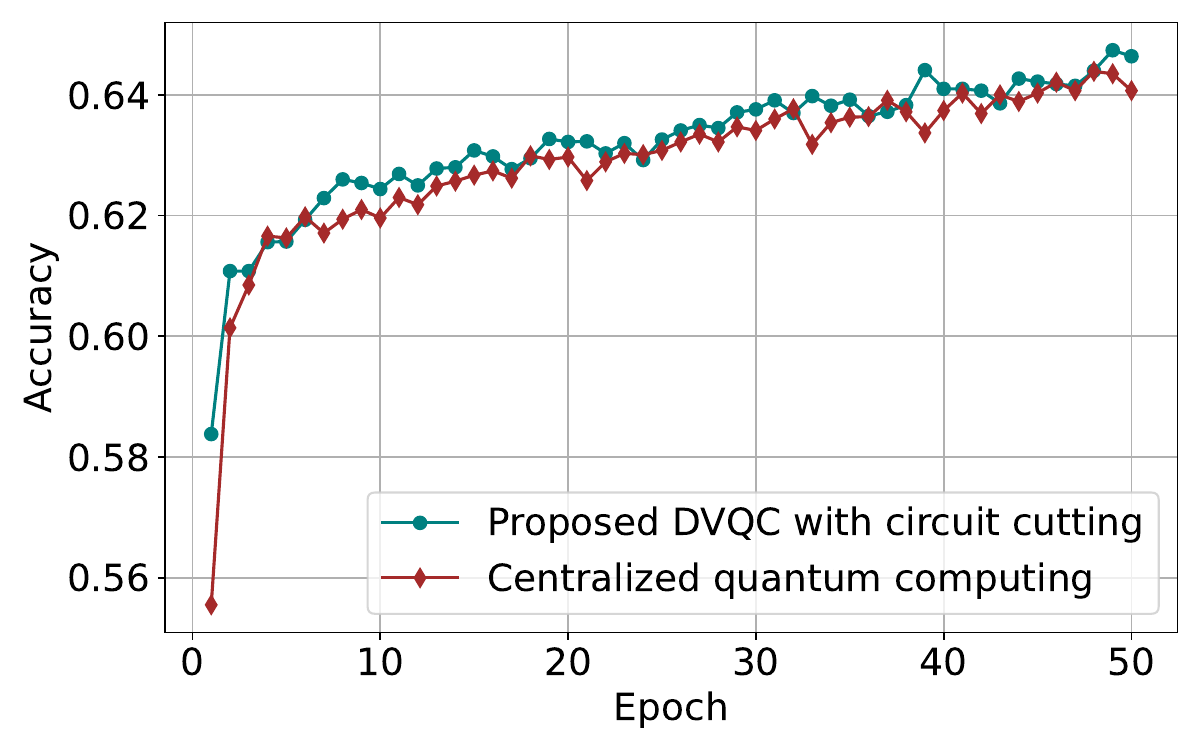}
        \caption{Training accuracy on MNIST dataset.}
        \label{learning11}
    \end{subfigure}
    \hfill 
    \begin{subfigure}[t]{0.49\linewidth} 
        \centering
        \includegraphics[width=\linewidth]{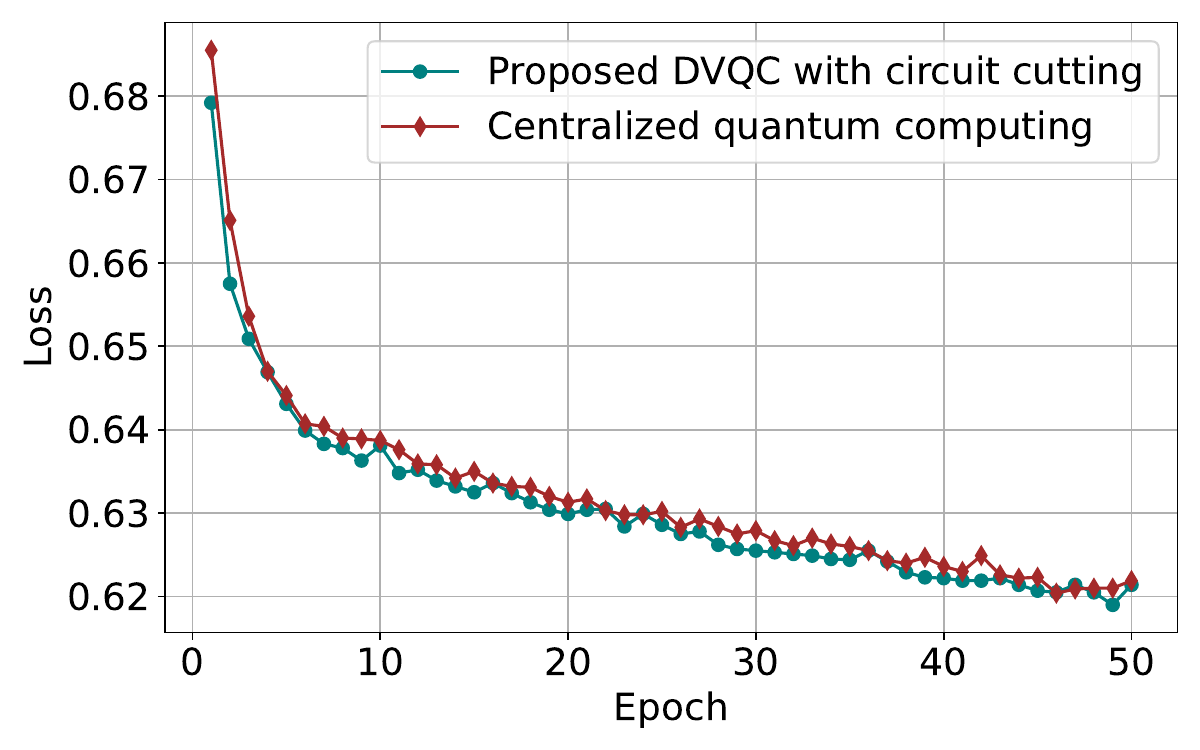}
        \caption{Training loss on MNIST dataset.}
        \label{learning12}
    \end{subfigure}
    \vspace{5pt} 
    \begin{subfigure}[t]{0.49\linewidth} 
        \centering
        \includegraphics[width=\linewidth]{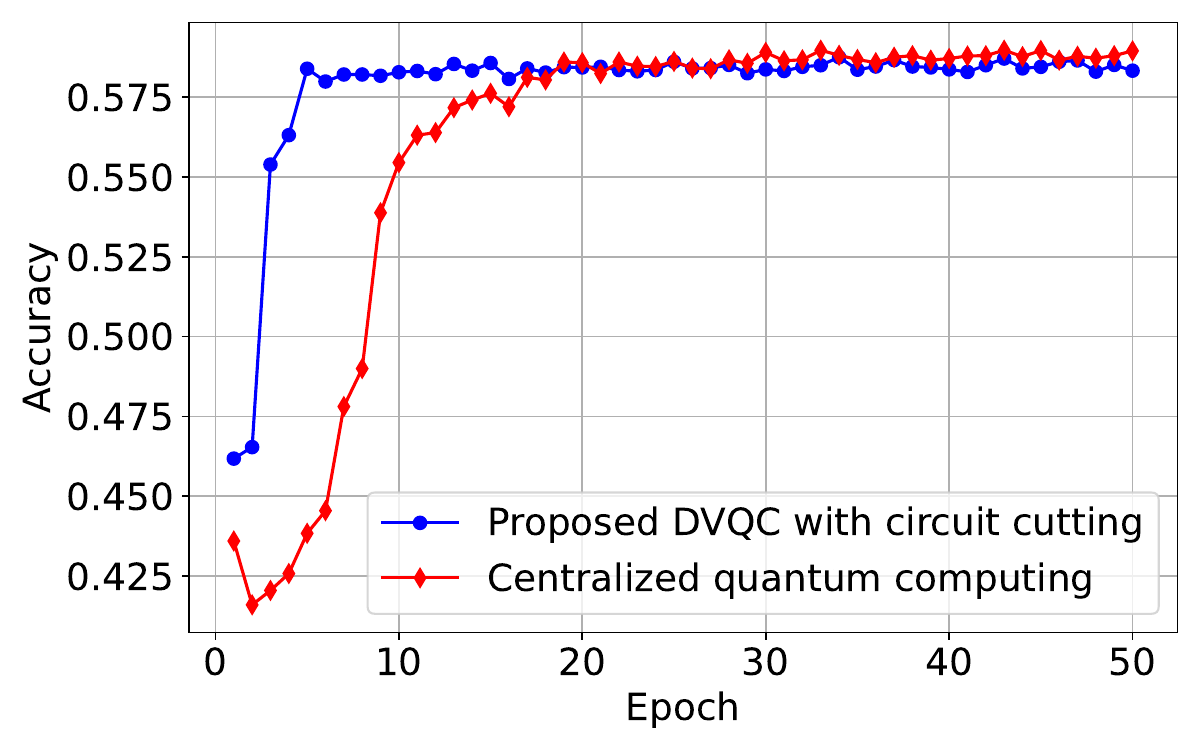}
        \caption{Training accuracy on CIFAR-10.}
        \label{learning13}
    \end{subfigure}
    \hfill 
    \begin{subfigure}[t]{0.49\linewidth} 
        \centering
        \includegraphics[width=\linewidth]{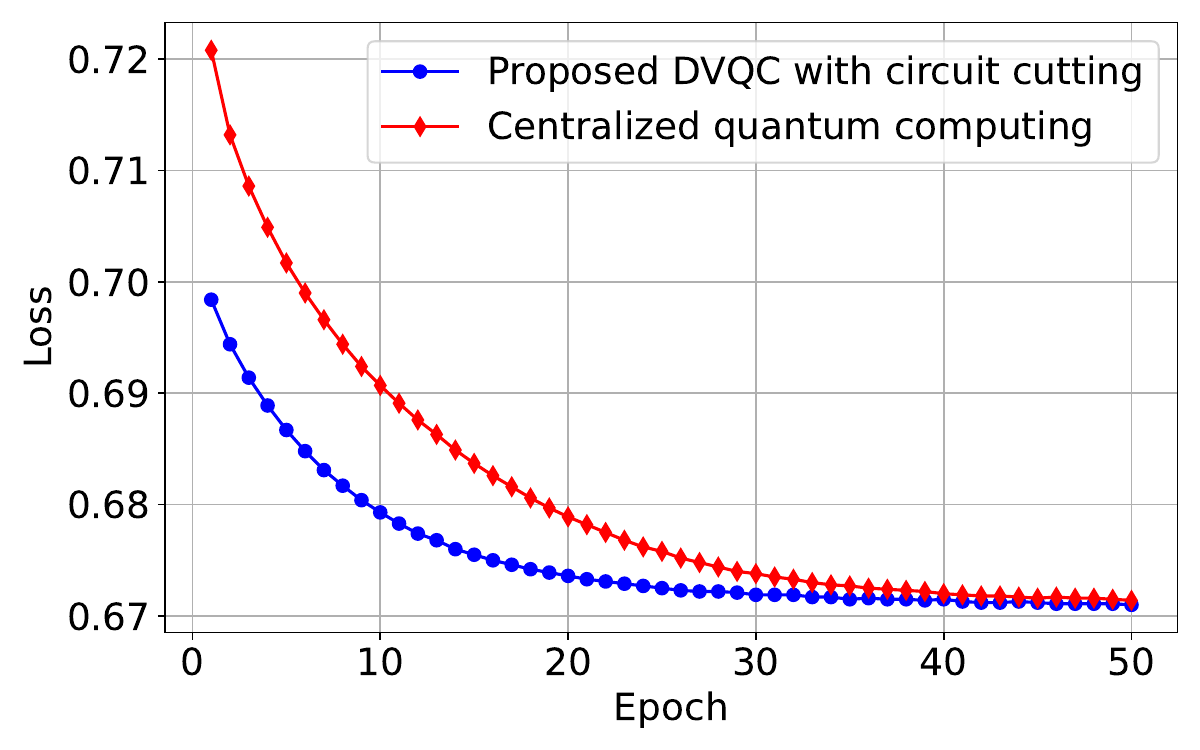}
        \caption{Training loss on CIFAR-10 dataset.}
        \label{learning14}
    \end{subfigure}
    \vspace{-3pt}
    \caption{Training accuracy and loss on MNIST and CIFAR-10 datasets using a 4-qubit PQC, where the proposed DVQC model applies circuit cutting by dividing the circuit into two 2-qubit subcircuits, compared against a centralized 4-qubit baseline.}
    \label{learning1}
\end{figure}

\subsection{Parameter Setting for DVQC Energy Minimization}
To evaluate the performance of our proposed energy-efficient DVQC framework, we conduct simulations that jointly optimize several critical parameters: processor’s PQC power consumption, the number of noise measurements, and resource allocation (communication transmission power). The simulation parameters are chosen to align with realistic operational settings \cite{yang2020energy,zeng2017energy}. All experiments are implemented in MATLAB using the YALMIP toolbox in conjunction with the MOSEK solver. The system bandwidth is considered to be of $20$ MHz. The processor’s total power budget for PQC operation and wireless transmission, denoted by $P_{k,\text{max}}^{(t)}$, ranges from $20$ to $40$ dB. The number of noise measurements ranges from $20$ to $40$. The noise power spectral density $n_{0}$ is assumed to be $-174$ dB.

\begin{figure}[h!]
    \centering
    \footnotesize
    \begin{subfigure}[t]{0.49\linewidth} 
        \centering
        \includegraphics[width=\linewidth]{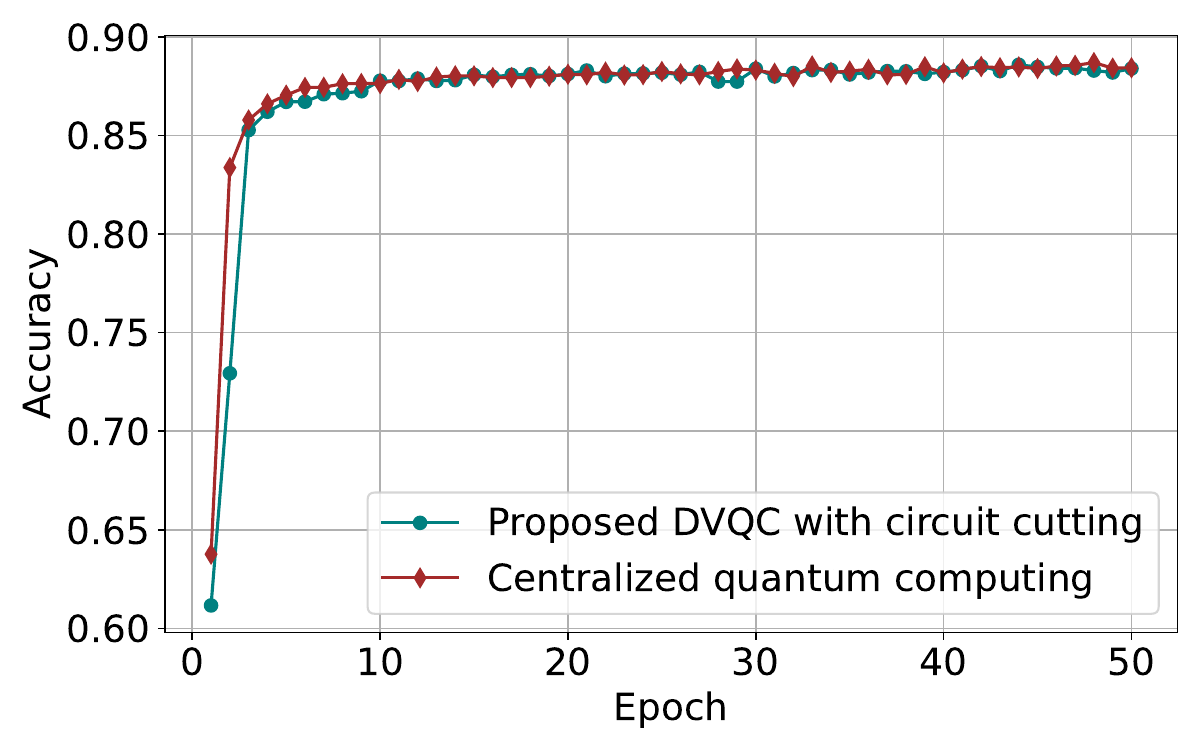}
        \caption{Training accuracy on MNIST dataset.}
        \label{fig4a}
    \end{subfigure}
    \hfill 
    \begin{subfigure}[t]{0.49\linewidth} 
        \centering
        \includegraphics[width=\linewidth]{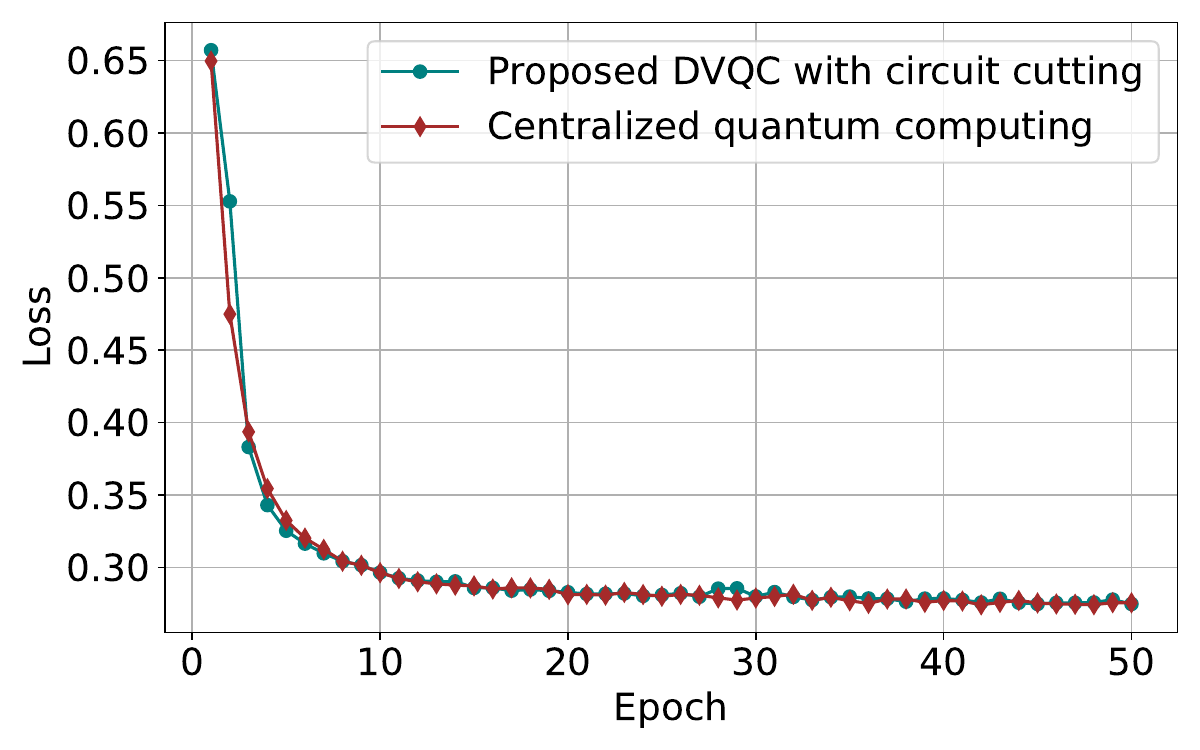}
        \caption{Training loss on MNIST dataset.}
        \label{fig4b}
    \end{subfigure}
    \vspace{5pt} 
    \begin{subfigure}[t]{0.49\linewidth} 
        \centering
        \includegraphics[width=\linewidth]{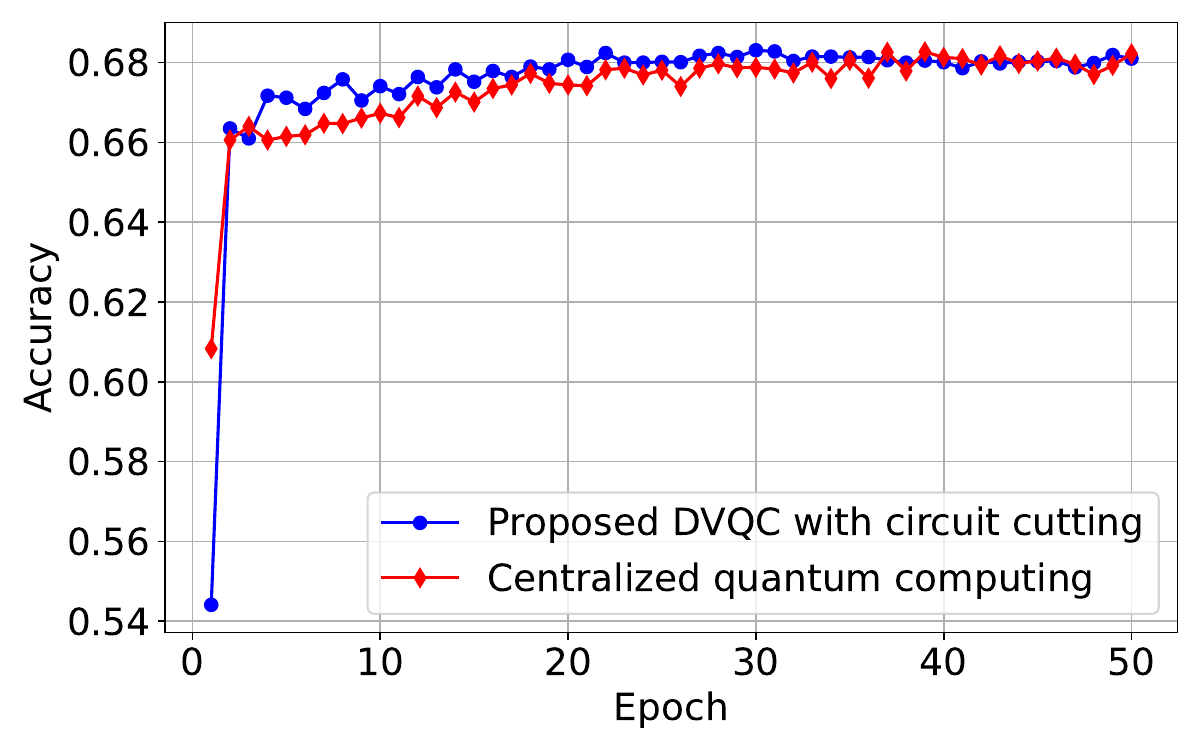}
        \caption{Training accuracy on CIFAR-10 dataset.}
        \label{fig4c}
    \end{subfigure}
    \hfill 
    \begin{subfigure}[t]{0.49\linewidth} 
        \centering
        \includegraphics[width=\linewidth]{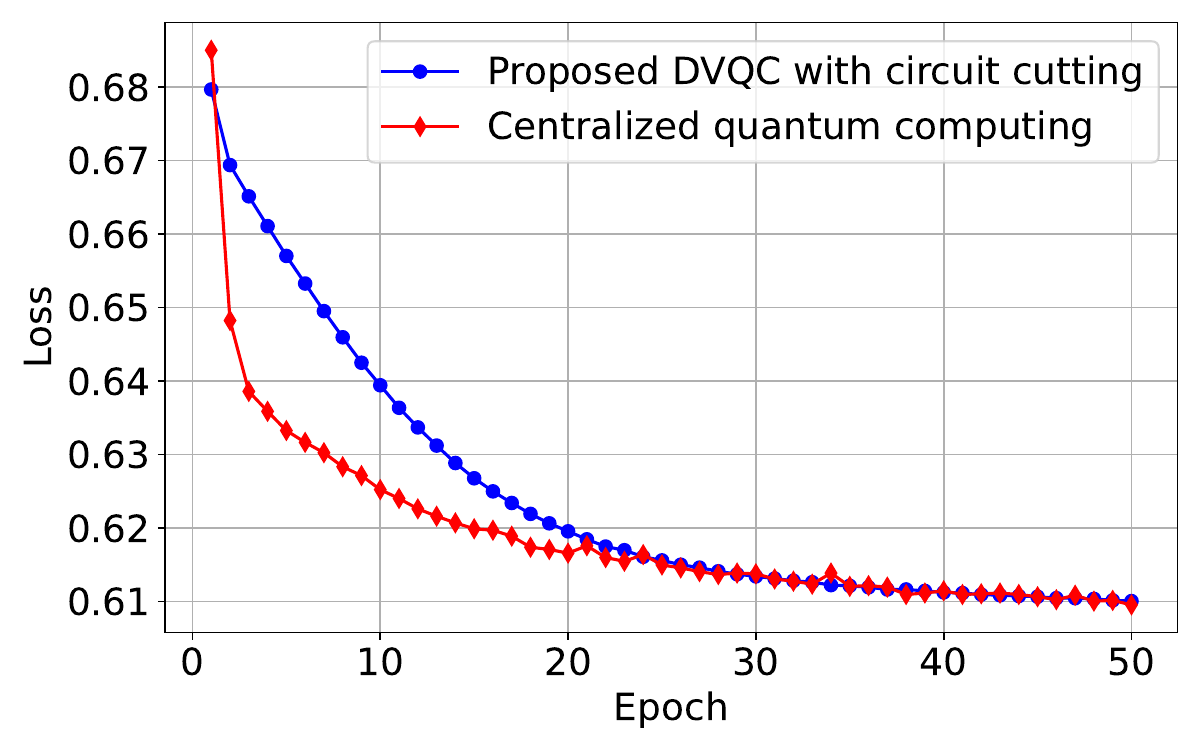}
        \caption{Training loss on CIFAR-10 dataset.}
        \label{fig4d}
    \end{subfigure}
    \vspace{-3pt}
    \caption{Training accuracy and loss on MNIST and CIFAR-10 datasets using a 6-qubit PQC, where the proposed DVQC model applies circuit cutting by splitting the circuit into three subcircuits consisting of 1, 2, and 3 qubits, respectively, compared against a centralized 6-qubit baseline.}
    \label{fig:2}
\end{figure}

\begin{figure}[h!]
    \centering
    \footnotesize
    \begin{subfigure}[t]{0.49\linewidth}
        \centering
        \includegraphics[width=\linewidth]{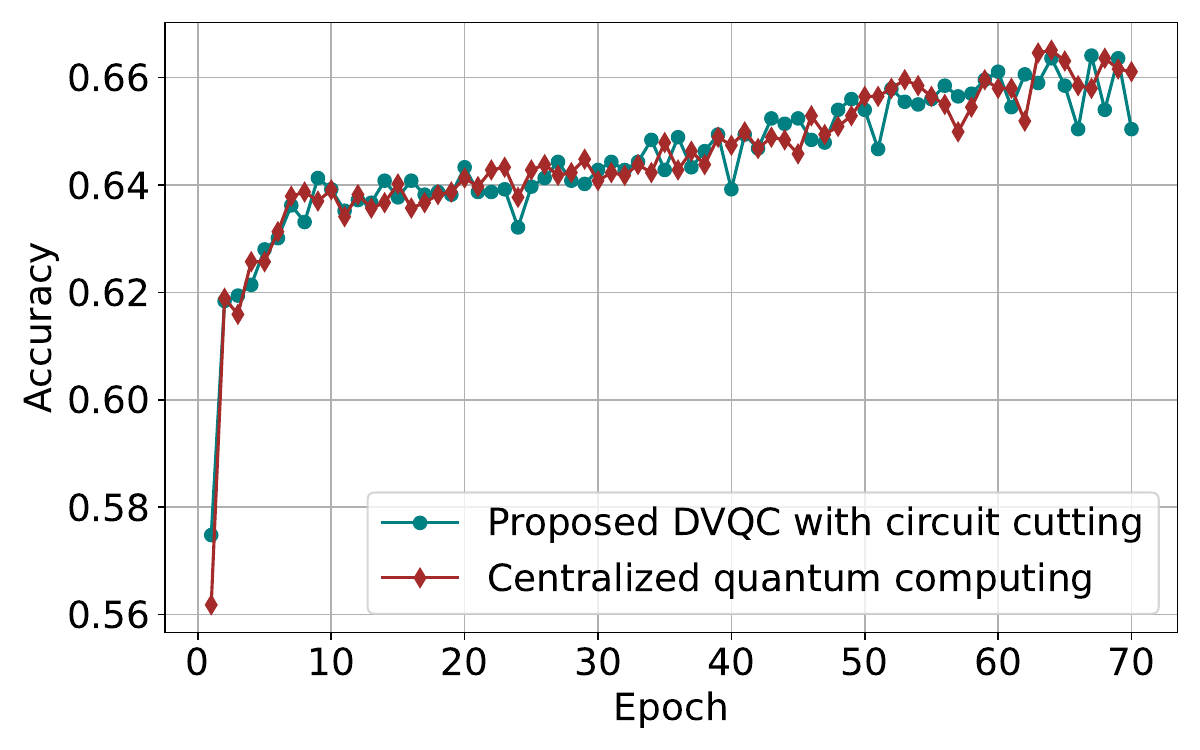}
        \caption{\textcolor{black}{Testing accuracy on MNIST dataset.}}
        \label{fig8a}
    \end{subfigure}
    \hfill 
    \begin{subfigure}[t]{0.49\linewidth} 
        \centering
        \includegraphics[width=\linewidth]{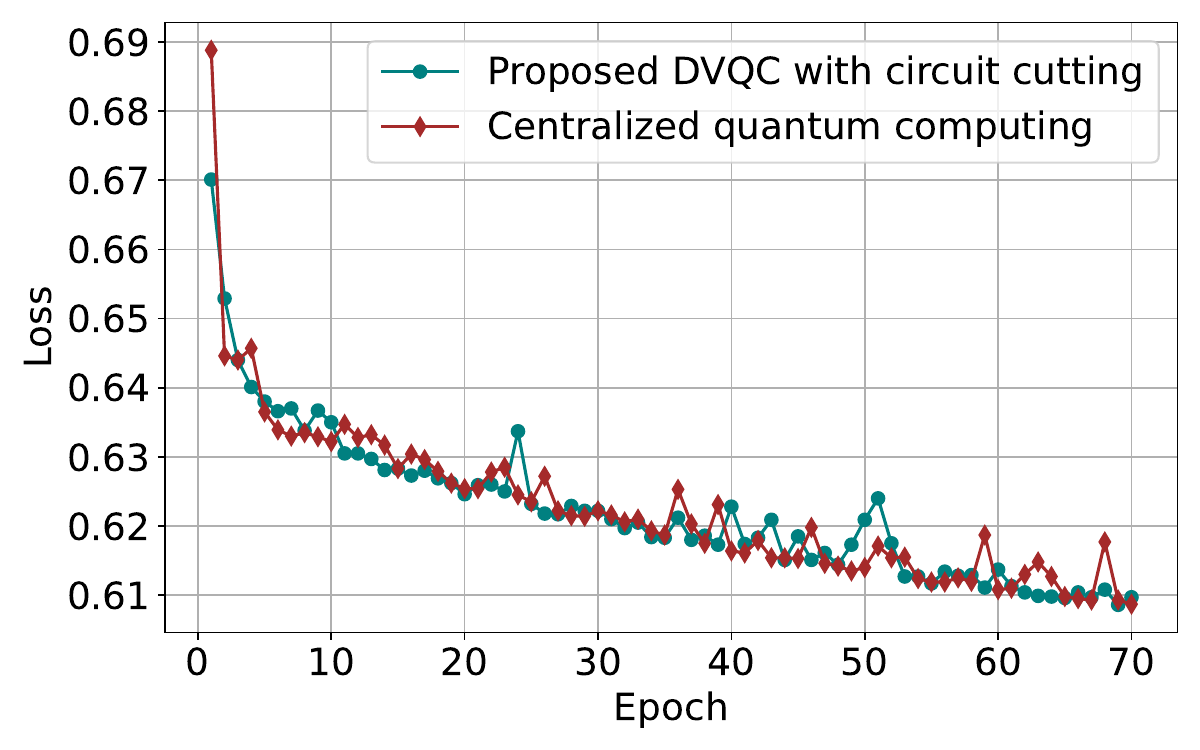}
        \caption{\textcolor{black}{Testing loss on MNIST dataset.}}
        \label{fig8b}
    \end{subfigure}
    \vspace{5pt}
    \begin{subfigure}[t]{0.49\linewidth}
        \centering
        \includegraphics[width=\linewidth]{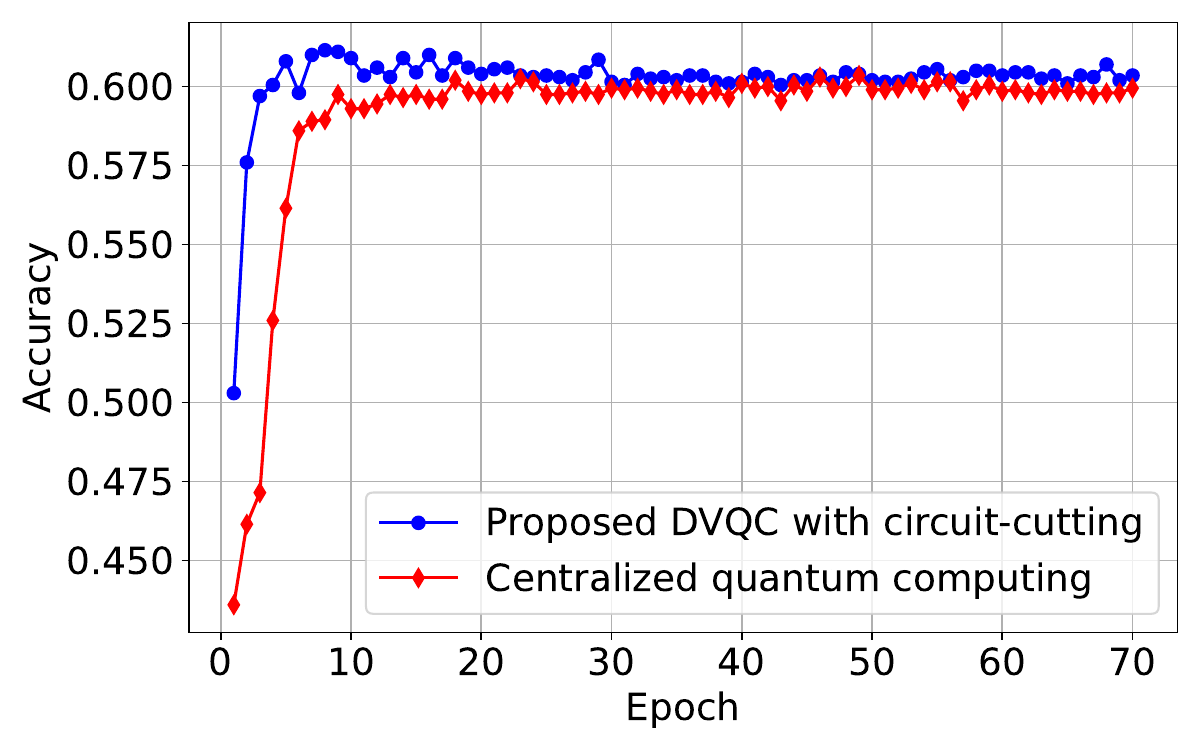}
        \caption{\textcolor{black}{Testing accuracy on CIFAR-10 dataset.}}
        \label{fig8c}
    \end{subfigure}
    \hfill
    \begin{subfigure}[t]{0.49\linewidth}
        \centering
        \includegraphics[width=\linewidth]{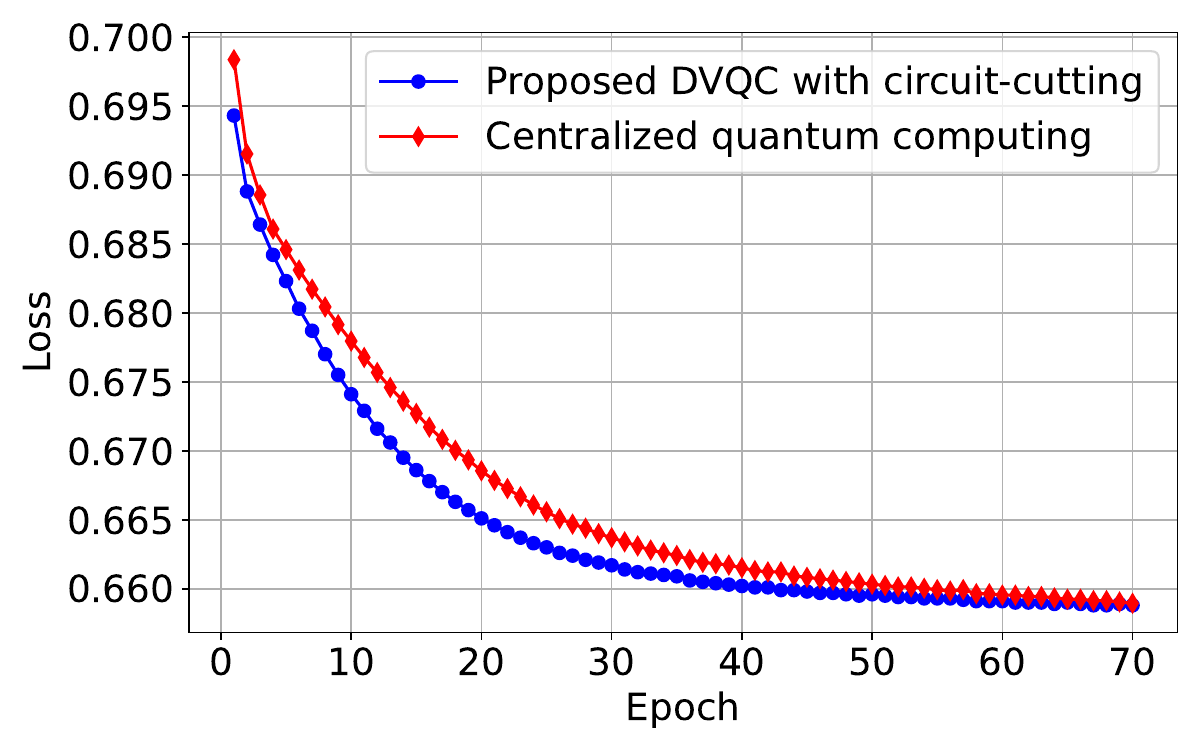}
        \caption{\textcolor{black}{Testing loss on CIFAR-10 dataset.}}
        \label{fig8d}
    \end{subfigure}
    \vspace{-3pt}
    \caption{\textcolor{black}{Testing accuracy and loss on MNIST and CIFAR-10 datasets using a 4-qubit PQC, where the proposed DVQC model applies circuit cutting by dividing the circuit into two 2-qubit subcircuits, compared against a centralized 4-qubit baseline.}}
    \label{fig:8}
\end{figure}

\begin{figure}[h!]
    \centering
    \footnotesize
    \begin{subfigure}[t]{0.49\linewidth}
        \centering
        \includegraphics[width=\linewidth]{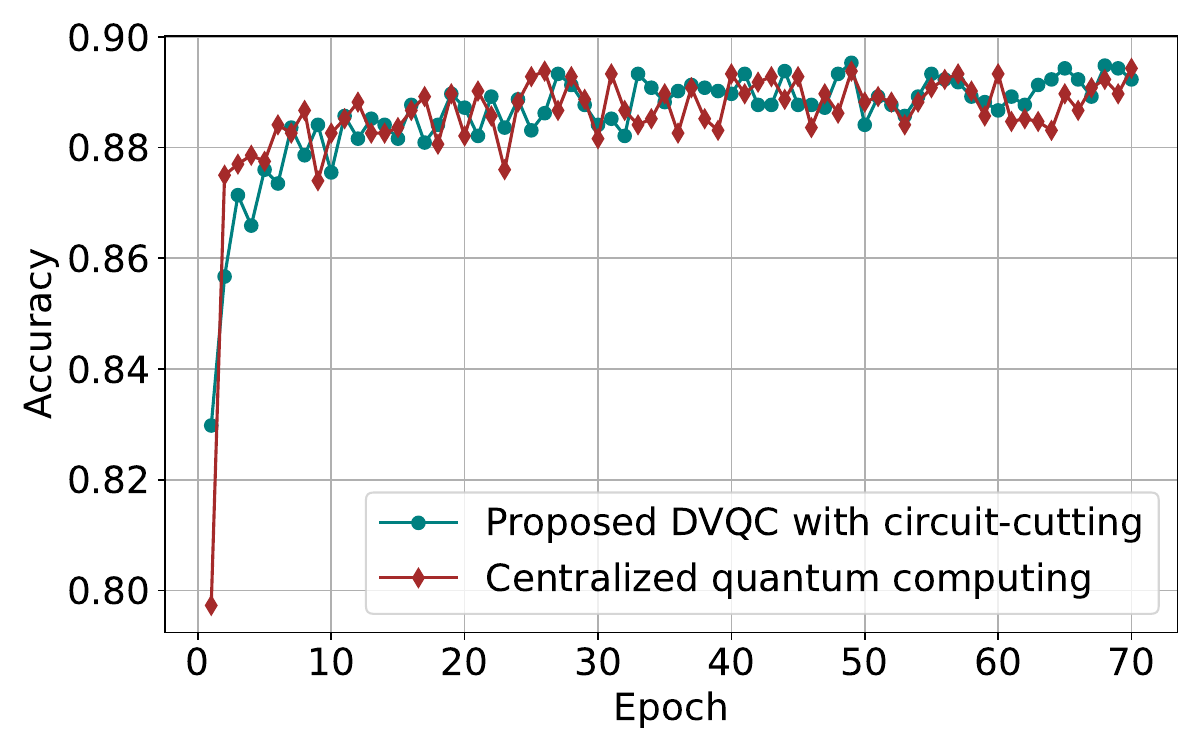}
        \caption{\textcolor{black}{Testing accuracy on MNIST dataset.}}
        \label{fig9a}
    \end{subfigure}
    \hfill 
    \begin{subfigure}[t]{0.49\linewidth} 
        \centering
        \includegraphics[width=\linewidth]{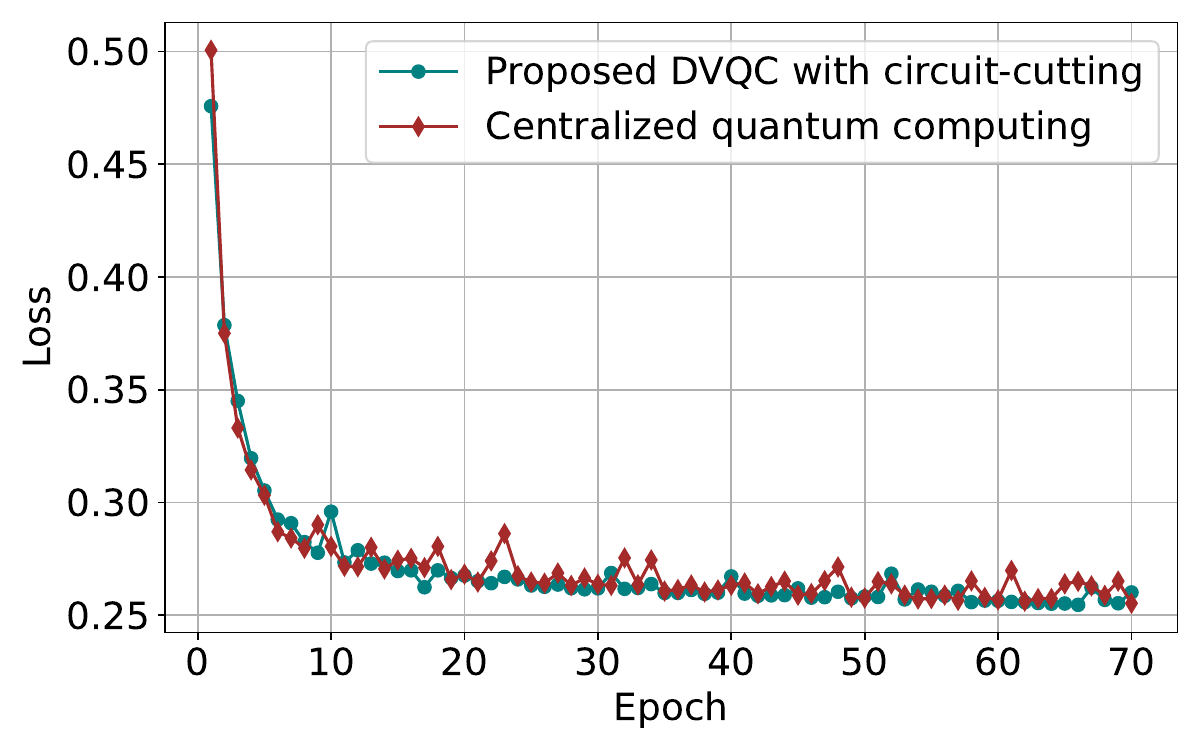}
        \caption{\textcolor{black}{Testing loss on MNIST dataset.}}
        \label{fig9b}
    \end{subfigure}
    \vspace{5pt}
    \begin{subfigure}[t]{0.49\linewidth}
        \centering
        \includegraphics[width=\linewidth]{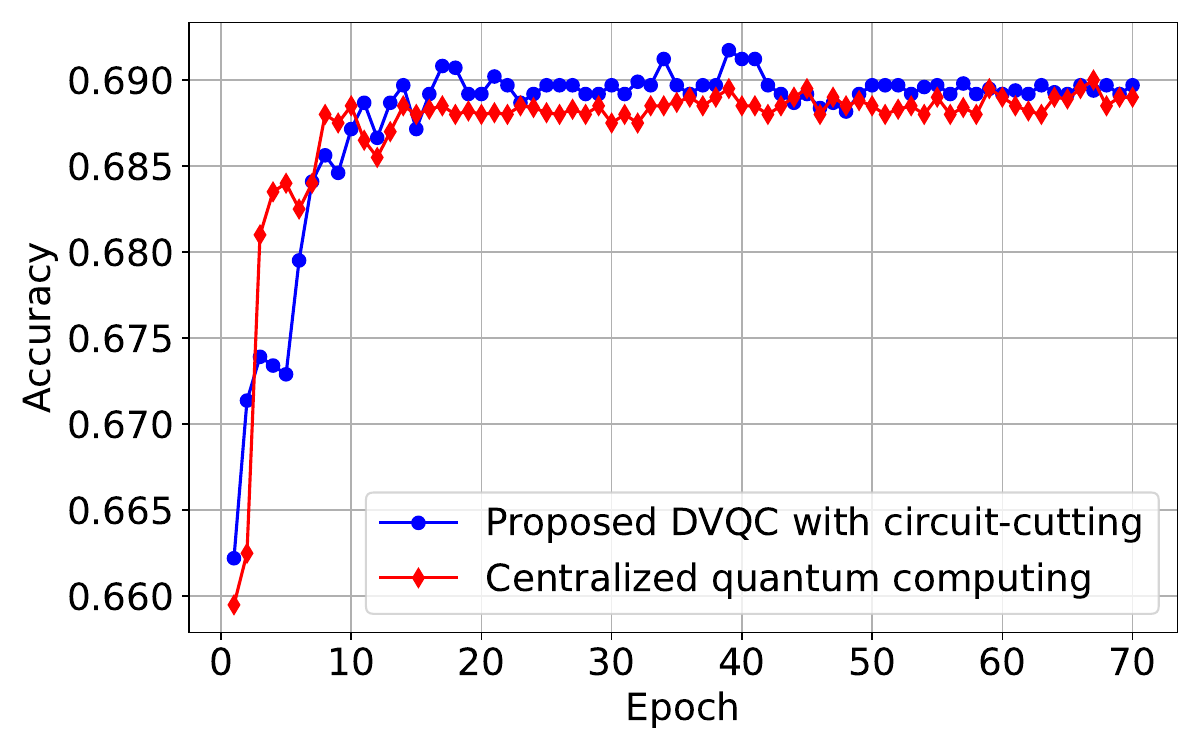}
        \caption{\textcolor{black}{Testing accuracy on CIFAR-10 dataset.}}
        \label{fig9c}
    \end{subfigure}
    \hfill
    \begin{subfigure}[t]{0.49\linewidth}
        \centering
        \includegraphics[width=\linewidth]{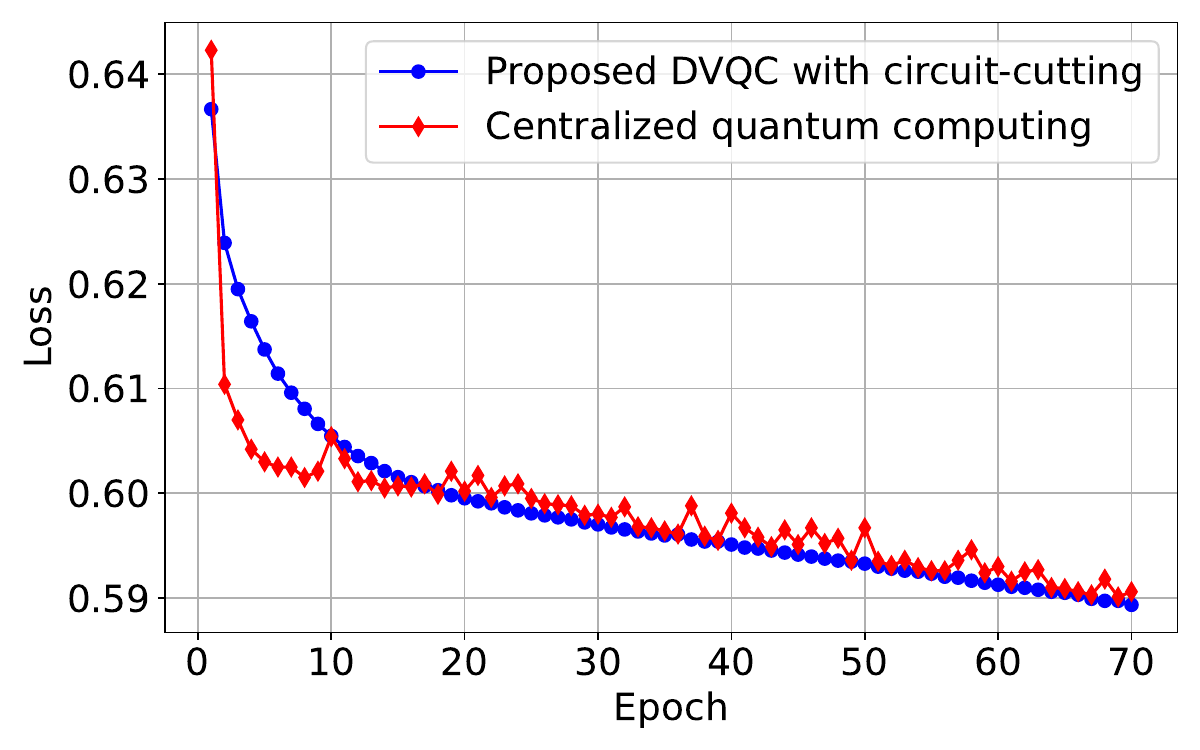}
        \caption{\textcolor{black}{Testing loss on CIFAR-10 dataset.}}
        \label{fig9d}
    \end{subfigure}
    \vspace{-3pt}
    \caption{\textcolor{black}{Testing accuracy and loss on MNIST and CIFAR-10 datasets using a 6-qubit PQC, where the proposed DVQC model applies circuit cutting by splitting the circuit into three subcircuits consisting of 1, 2, and 3 qubits, respectively, compared against a centralized 6-qubit baseline.}}
    \label{fig:9'}
\end{figure}

\begin{figure}[h!]
    \centering
    \includegraphics[width=1.1\linewidth]{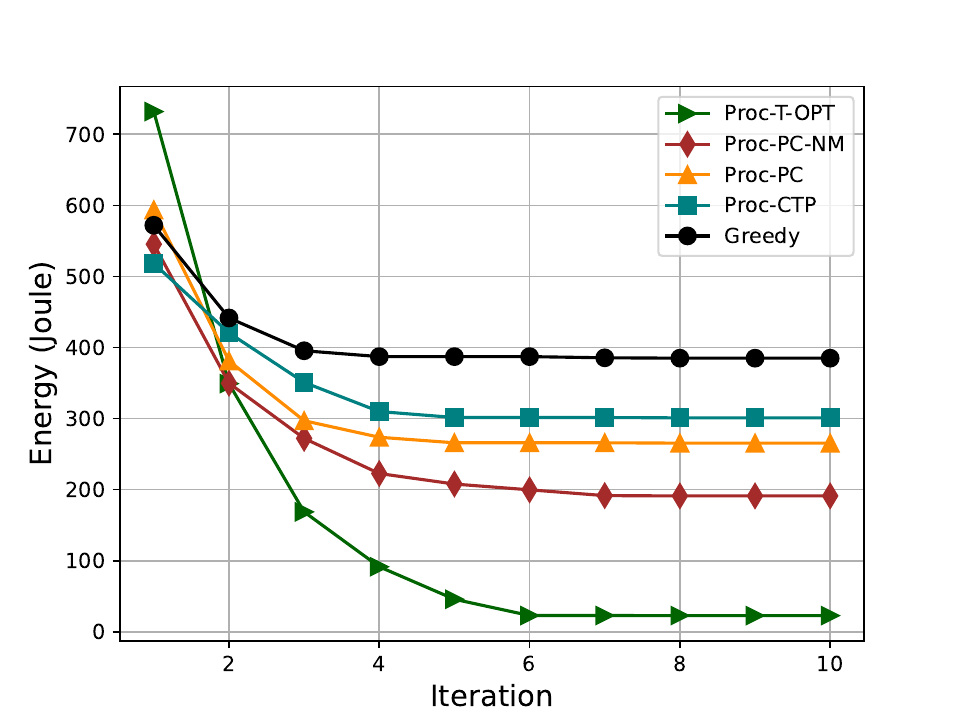} 
    \caption{Comparison of energy optimization performance among different schemes.}
    \label{fig:5}  
\end{figure}

\begin{figure}[h!]
    \centering
    \begin{subfigure}[b]{0.49\linewidth}
        \centering
        \includegraphics[width=\linewidth]{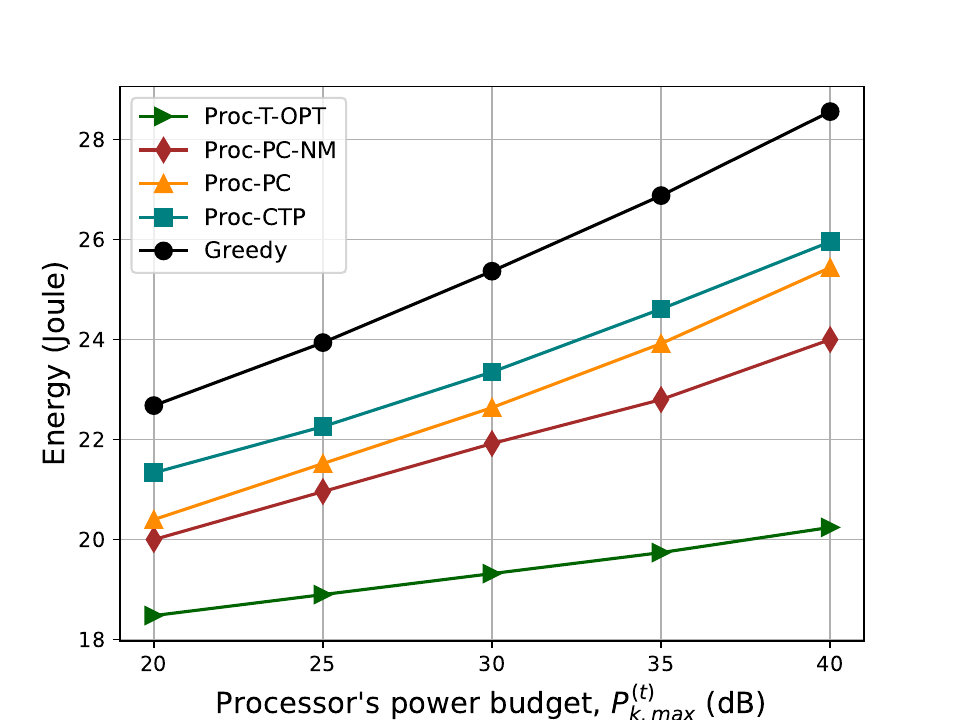}
        \caption{Processor's energy consumption versus its maximum allowable power threshold.}
        \label{fig:6}
    \end{subfigure}
    \hfill
    \begin{subfigure}[b]{0.49\linewidth}
        \centering
        \includegraphics[width=\linewidth]{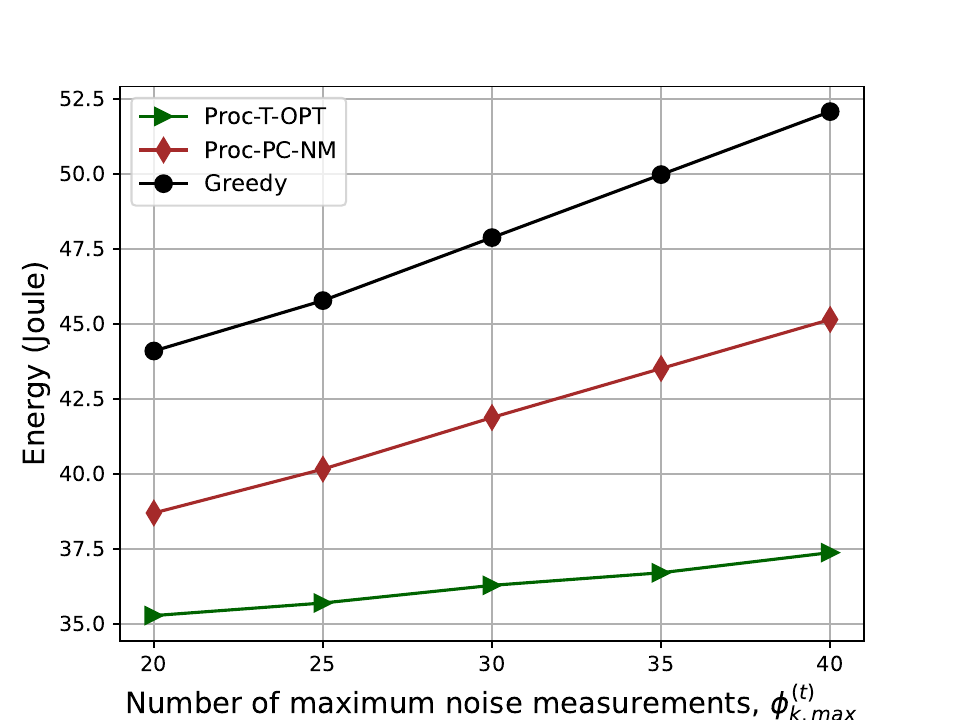}
        \caption{Processor's energy consumption versus its maximum noise measurements.}
        \label{fig:7}
    \end{subfigure}
    \caption{Performance evaluation of processor's energy consumption under varying system parameters.}
    \label{fig:7combined}
\end{figure}

\begin{figure*}[h!]
    \centering
    \begin{subfigure}{0.32\textwidth}
        \centering
        \includegraphics[width=\linewidth]{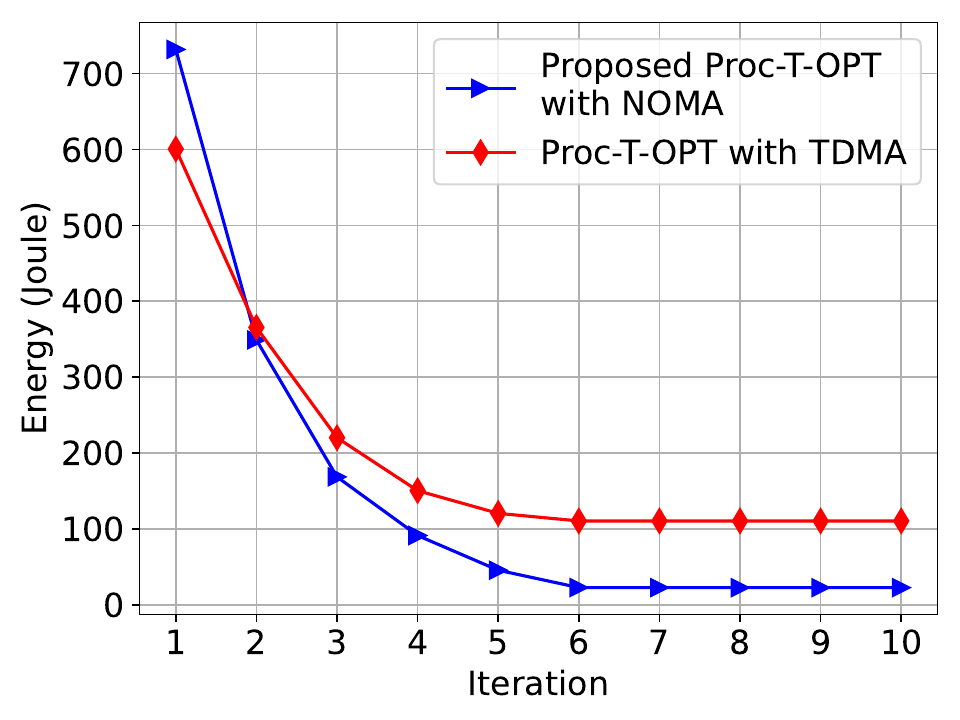} 
        \caption{\footnotesize \textcolor{black}{Convergence of the energy minimization approach Proc-T-OPT under NOMA and TDMA transmission schemes.}}
        \label{fig11}
    \end{subfigure}
    \hfill
    \begin{subfigure}{0.32\textwidth}
        \centering
        \includegraphics[width=\linewidth]{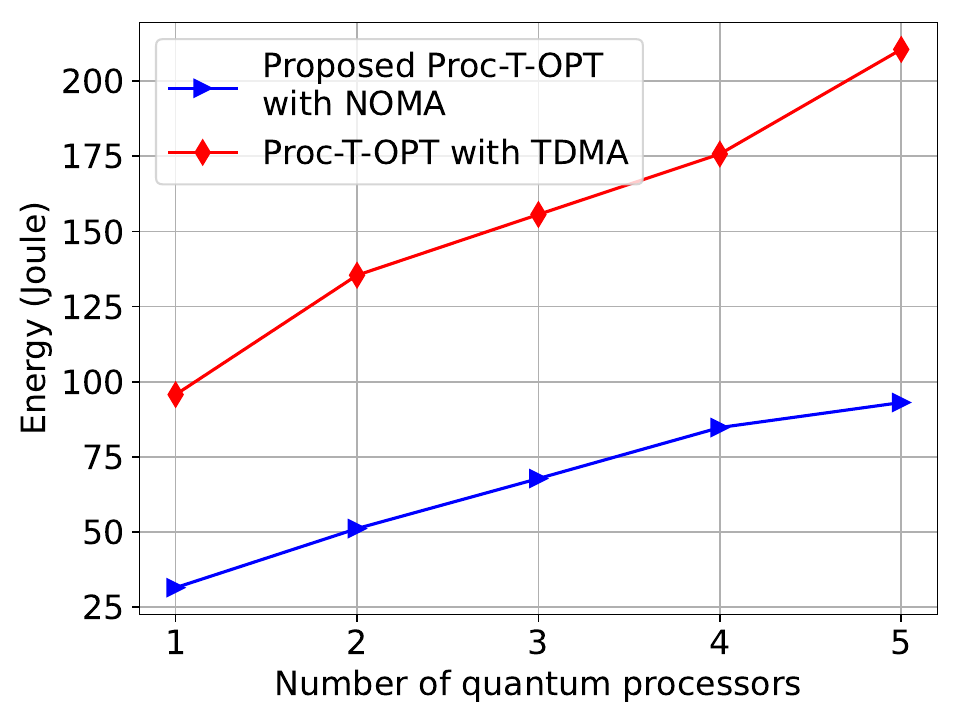} 
        \caption{\footnotesize \textcolor{black}{Processor energy consumption versus the number of selected quantum processors under NOMA and TDMA.}}
        \label{fig12}
    \end{subfigure}
    \hfill
    \begin{subfigure}{0.32\textwidth}
        \centering
        \includegraphics[width=\linewidth]{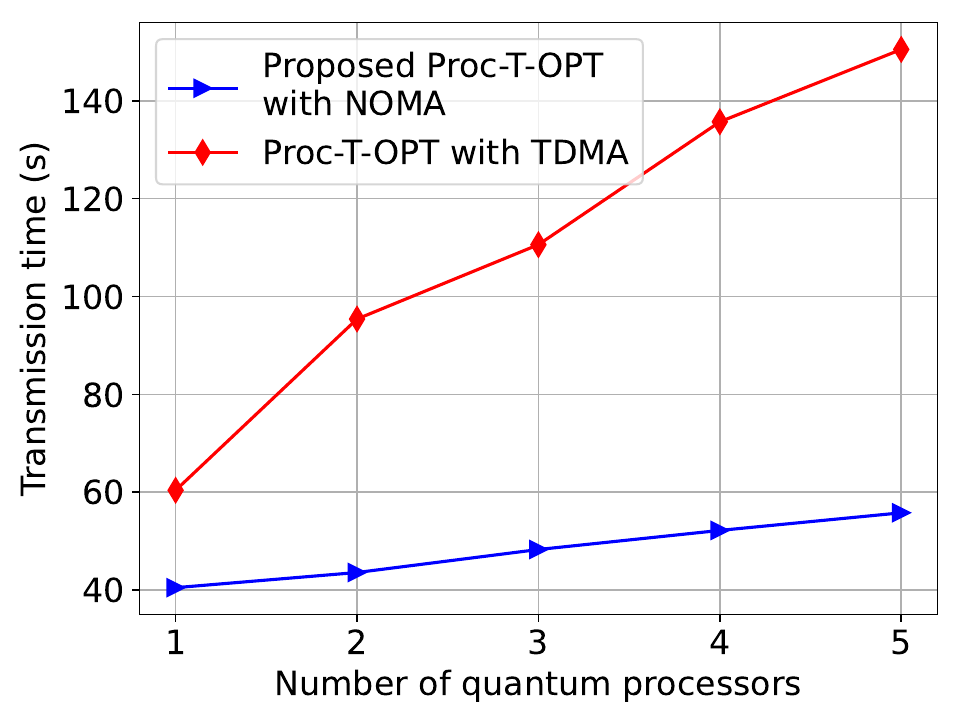}
        \caption{\footnotesize \textcolor{black}{Measurement outcome transmission time versus the number of selected quantum processors under NOMA and TDMA.}}
        \label{fig13}
    \end{subfigure}
    \caption{\footnotesize \textcolor{black}{Comparison between the proposed NOMA-based DVQC framework and the TDMA baseline in terms of overall energy consumption and transmission performance.}}
    \label{fig14}
\end{figure*}

\subsection{Parameter Setting for Utility Maximization for Incentive Mechanism}
In our utility maximization experiments, we consider a scenario with $K=10$ candidate quantum processors. The payment offer $m_{k}$ lies within the range of 10 to 50 units, reflecting the monetary compensation proposed by the aggregator for executing the assigned subcircuit and transmitting measurement outcomes. The energy cost $E_{k}^{(t)}$ ranges from 10 to 30 Joules, capturing both the quantum computation energy and the communication energy required by processor. The qubit capacity $q_{k}$ varies between 2 and 7 qubits, indicating the number of qubits available for subcircuit execution on each processor. The total incentive budget at the aggregator is set to $M_{\text{max}}=120$ units, and the required qubit coverage to complete the centralized PQC is $Q=12$. The aggregator solves a utility maximization problem under these constraints using the ADMM algorithm, aiming to select a subset of processors that maximizes the system utility while satisfying the budget and qubit constraints.

\subsection{DVQC Model Training Performance}

Fig.~\ref{learning1} shows training accuracy and loss performance on the MNIST and CIFAR-10 datasets using a 4-qubit PQC, where the proposed DVQC model applies circuit cutting by dividing the circuit into two 2-qubit subcircuits, each assigned to a separate quantum processor, and compares it against a 4-qubit centralized quantum computing baseline. Fig.~\ref{learning11} presents the training accuracy on the MNIST dataset where both models exhibit competitive performance, maintaining similar accuracy levels across 50 training epochs. This demonstrates that DVQC can effectively scale quantum computing models without sacrificing learning performance, making it a viable solution for near-term, resource-constrained quantum hardware. Fig.~\ref{learning12} illustrates the corresponding training loss where both approaches demonstrate a consistent downward trend. The DVQC model, which utilizes two 2-qubit subcircuits, closely tracks the centralized model's loss performance throughout the epochs. Minor fluctuations in the centralized model suggest slightly higher sensitivity to noise. Fig.~\ref{learning13} demonstrates the training accuracy on the CIFAR-10 dataset under similar settings, exhibiting a trend consistent with the MNIST results. Fig.~\ref{learning14} presents the corresponding training loss, where DVQC and centralized models exhibit similar behavior, confirming that DVQC maintains stable and effective learning even for more complex datasets. 

Fig.~\ref{fig:2} shows the training accuracy and loss on the MNIST (Fig.~\ref{fig4a} and Fig.~\ref{fig4b}) and CIFAR-10 (Fig.~\ref{fig4c} and Fig.~\ref{fig4d}) datasets using a 6-qubit PQC. The proposed DVQC model applies circuit cutting by splitting the 6-qubit PQC into three smaller subcircuits of 1, 2, and 3 qubits. It achieves similar \textcolor{black}{final training accuracy and loss} to the centralized 6-qubit model on both MNIST and CIFAR-10 datasets. \textcolor{black}{Although the proposed DVQC framework exhibits convergence behavior similar to centralized quantum computing on MNIST, it shows slower initial convergence on the more complex CIFAR-10 dataset, as observed in Fig.~\ref{fig4d}. Initially, the proposed DVQC framework requires additional training epochs to obtain the loss performance achieved by centralized quantum computing. The slower initial convergence is attributed to the partitioning of the original PQC into smaller subcircuits, which changes the circuit architecture and parameterization, potentially affecting gradient-based optimization. This effect may be more pronounced for complex datasets such as CIFAR-10, despite the coordinated parameter updates guided by a common global loss. Nevertheless, the difference in training loss gradually diminishes, and both approaches achieve approximately the same final loss after 50 epochs. These results demonstrate that by distributing the computation across smaller subcircuits, DVQC achieves comparable final training performance, despite differences in convergence speed, making it practical for near-term quantum hardware with limited qubit capacity.} We note that the objective of DVQC is not to surpass centralized quantum computing. Rather, it aims to enable feasible execution of large PQCs under realistic qubit limitations by distributing the workload across multiple quantum processors with smaller qubit capacities, while still achieving comparable performance \textcolor{black}{in terms of final accuracy and loss.} Therefore, these results are consistent with the theoretical design of the proposed DVQC  framework. Specifically, although the original PQC is partitioned into multiple smaller subcircuits, all processors are jointly optimized through a common global loss computed at the aggregator, enabling coordinated parameter updates across the distributed quantum processors. \textcolor{black}{Consequently, circuit cutting primarily reduces the per-processor qubit requirement while allowing the distributed framework to achieve similar final learning performance, although its convergence rate may differ from that of centralized quantum computing, particularly for more complex datasets such as CIFAR-10.} These observations validate that the proposed DVQC framework effectively addresses the practical qubit limitations without introducing degradation in learning capability.

\textcolor{black}{To further evaluate the generalization capability of the proposed DVQC framework, Fig.~\ref{fig:8} and Fig.~\ref{fig:9'} present the testing accuracy and testing loss on the MNIST and CIFAR-10 datasets under the 4-qubit (2+2) and 6-qubit (1+2+3) PQC partitioning configurations, respectively, compared with their corresponding centralized quantum computing baselines. As shown in Fig.~\ref{fig8a} and Fig.~\ref{fig8b}, the proposed DVQC framework achieves testing accuracy and loss comparable to the centralized 4-qubit model on MNIST, with both approaches exhibiting similar performance trends. Likewise, Fig.~\ref{fig8c} and Fig.~\ref{fig8d} demonstrate that DVQC closely matches the testing accuracy and loss of centralized quantum computing on CIFAR-10, while exhibiting slightly faster initial improvement in accuracy and reduction in loss. For the 6-qubit configuration, Fig.~\ref{fig9a} and Fig.~\ref{fig9b} show that the proposed DVQC and centralized quantum computing models achieve nearly identical testing accuracy and loss on MNIST. Meanwhile, Fig.~\ref{fig9c} and Fig.~\ref{fig9d} indicate that the DVQC achieves similar testing accuracy to centralized quantum computing on CIFAR-10, although its testing loss converges more slowly during the initial epochs, with both approaches eventually reaching similar loss values. Overall, these testing results complement the training performance and demonstrate that the proposed DVQC framework achieves testing accuracy and loss comparable to centralized quantum computing across both datasets and circuit partitioning configurations, providing further evidence of its ability to generalize to previously unseen data while reducing the qubit requirements of individual quantum processors.}

\subsection{DVQC Energy Optimization Performance}
We compare our proposed joint optimization approach, referred to as Proc-T-OPT, with four benchmark schemes: (i) Proc-PC-NM, which optimizes only processor's PQC power control and number of noise measurements; (ii) Proc-PC, which optimizes only processor's PQC power control, (iii) Proc-CTP, which optimizes only processor's communication transmit power; and (iv) a Greedy scheme that performs uncoordinated resource allocation without joint consideration of optimization variables. Fig.~\ref{fig:5} compares the energy optimization performance of Proc-T-OPT with four benchmark schemes over ten iterations. As illustrated in the figure, Proc-T-OPT achieves substantial reduction in processor energy consumption, converging to a much lower energy level than the other schemes. After six iterations, the energy consumed by Proc-T-OPT is reduced by approximately $89.74\%$, $92.59\%$, $93.33\%$, and $94.87\%$ relative to Proc-PC-NM, Proc-PC, Proc-CTP, and Greedy, respectively. \textcolor{black}{The superior performance of Proc-T-OPT stems from the joint optimization of quantum computation and communication resources through the proposed BCD-SCA framework. By jointly optimizing PQC gate powers, number of noise measurements, and transmit power, the algorithm effectively exploits the interdependence among these variables, leading to a significantly lower overall energy consumption than optimizing only a subset of variables.}

Fig.~\ref{fig:6} shows how processor energy consumption varies with its maximum allowable power budget for different optimization schemes. As the power budget increases, energy consumption rises for all methods. However, Proc-T-OPT consistently achieves the lowest energy consumption across all power budget levels. In contrast, Greedy and Proc-CTP incur significantly higher energy use, while Proc-PC-NM and Proc-PC provide moderate improvements. These results highlight that joint optimization in Proc-T-OPT leads to more efficient energy use, especially as resource constraints are relaxed. Fig.~\ref{fig:7} illustrates the effect of increasing the maximum number of noise measurements on processor energy consumption. Across all schemes, energy consumption increases with more noise measurements, but the rate of increase is much lower for Proc-T-OPT compared to the others. Although Proc-PC-NM outperforms Greedy, it lags behind Proc-T-OPT, emphasizing the advantage of comprehensive joint optimization over partial or uncoordinated strategies. This demonstrates Proc-T-OPT’s effectiveness in controlling energy costs even as noise measurement demands grow. \textcolor{black}{The observed trends are consistent with the proposed processor energy model, where both the maximum budget and the number of noise measurements directly influence the total energy consumption. A larger power budget allows higher operating power during quantum gate execution and communication, while increasing the number of noise measurements requires additional circuit executions, thereby increasing quantum computing energy. Since Proc-T-OPT jointly optimizes all energy-related variables, it results in significantly slower increase in energy consumption than the benchmark schemes as system resources increase.} \textcolor{black}{We note that only Proc-T-OPT, Proc-PC-NM, and the Greedy scheme are shown in Fig.~\ref{fig:7}, since the maximum number of noise measurements $\phi_{k,\text{max}}^{(t)}$ affects only schemes that optimize number of noise measurements $\phi_{k}^{(t)}$, whereas Proc-PC and Proc-CTP are independent of this parameter.}

\textcolor{black}{Fig.~\ref{fig14} compares the proposed NOMA-based DVQC framework with the TDMA baseline in terms of energy consumption and transmission performance. Fig.~\ref{fig11} demonstrates the convergence performance of the joint optimization approach Proc-T-OPT for processor energy minimization under NOMA and TDMA schemes. \textcolor{black}{It is observed that the processor energy consumption decreases sharply during the first few iterations before reaching a stable minimum, while consistently achieving lower energy under NOMA than under TDMA. This behavior directly follows from the proposed NOMA-based uplink model, where simultaneous measurement transmission reduces idle waiting, thereby shortening the communication phase of each DVQC iteration. Consequently, the proposed BCD-SCA optimization converges to a lower operating point because the transmission energy component in the processor energy model is significantly reduced compared with the TDMA scheme.} TDMA enforces sequential time-slot allocation for each processor, leading to prolonged transmission duration due to increased waiting and idle periods, which in turn amplifies cumulative communication energy consumption. Fig.~\ref{fig12} shows the average processor energy consumption versus the number of selected processors. As more processors are selected, the energy under TDMA increases sharply due to sequential time-slot allocation, which introduces long idle and waiting periods where processors remain active and consume additional energy. In contrast, NOMA allows simultaneous transmission of measurement outcomes, significantly reducing waiting time and communication duration. However, the average energy consumption under NOMA still increases slightly with the number of processors, as the presence of slower processors can extend the effective transmission phase. Fig.~\ref{fig13} presents the total transmission time versus the number of selected processors, where TDMA exhibits significantly higher latency because processors transmit sequentially in orthogonal time slots, while NOMA enables concurrent transmissions using power-domain multiplexing and SIC at the aggregator, resulting in substantially reduced transmission delay.} \textcolor{black}{These results validate the design choice of incorporating a NOMA-based uplink into the proposed DVQC framework. Since measurement transmission is performed after every training iteration, reducing the communication latency also shortens the overall DVQC cycle, thereby improving the practicality and scalability of the proposed framework.} 

\begin{figure}[t!]
    \centering
    \includegraphics[width=0.99\linewidth]{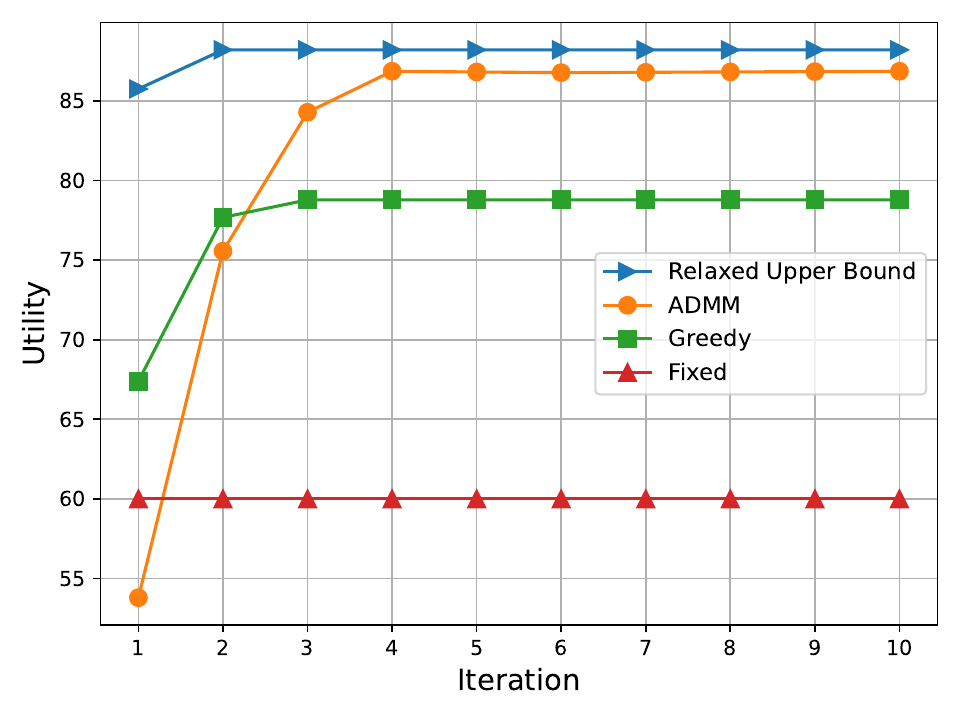} 
    \caption{Convergence of utility for different selection algorithms over iterations.}
    \label{fig:energy4schemes}  
\end{figure}

\begin{figure}[t!]
    \centering
    \begin{subfigure}[b]{0.49\linewidth}
        \centering
        \includegraphics[width=\linewidth]{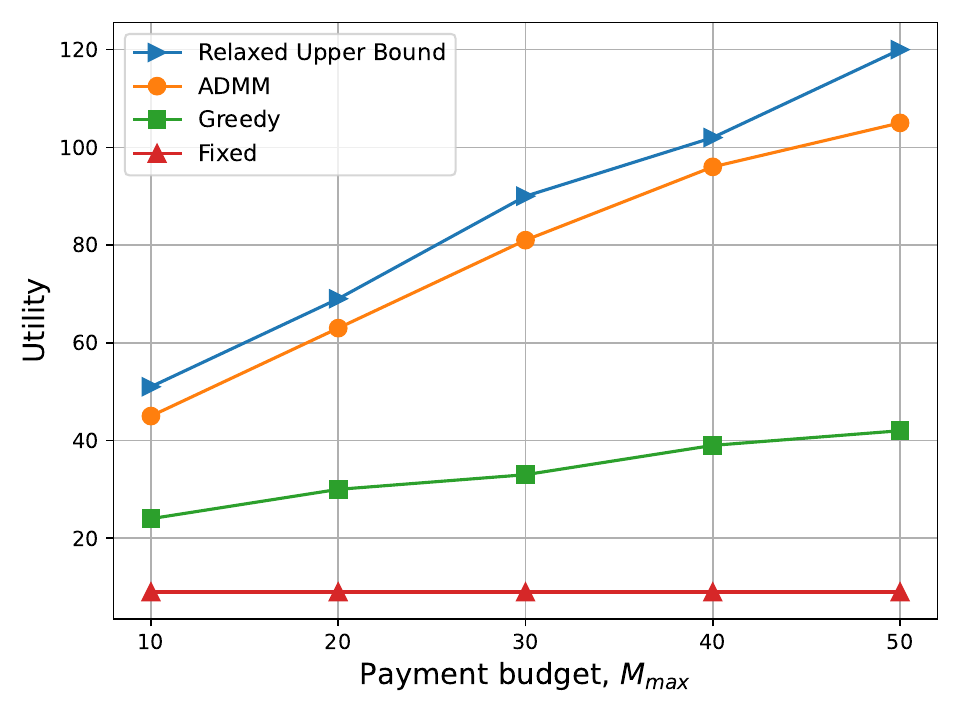}
        \caption{Utility versus payment budget, showing how increased budget improves achievable utility.}
        \label{fig:9}
    \end{subfigure}
    \hfill
    \begin{subfigure}[b]{0.49\linewidth}
        \centering
        \includegraphics[width=\linewidth]{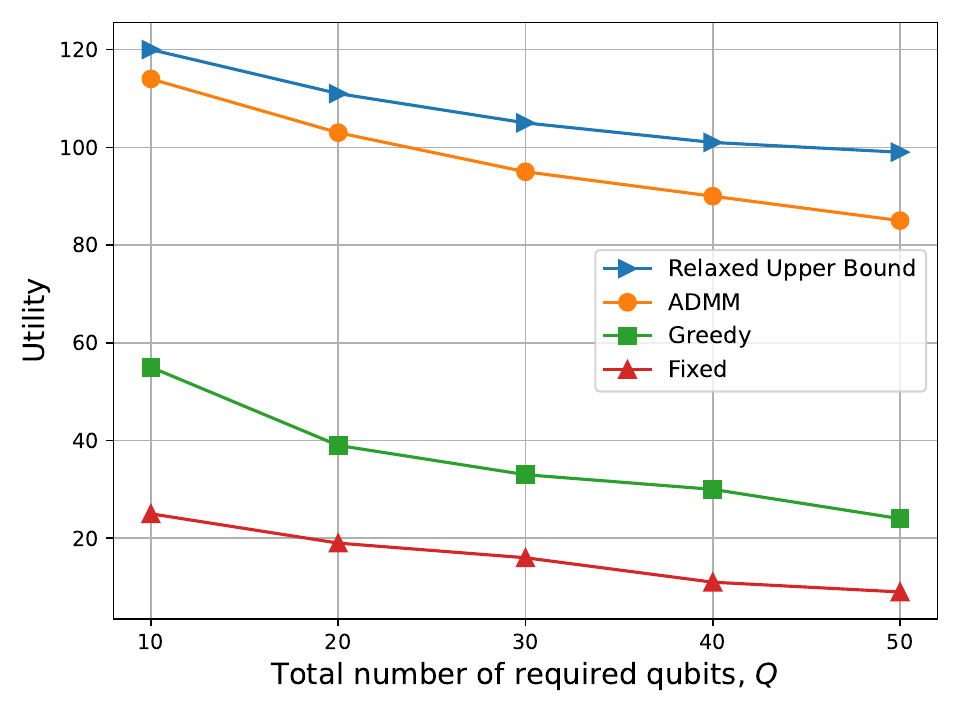}
        \caption{ Utility versus required qubits, illustrating the impact of increased qubit coverage demands on system utility.}
        \label{fig:10}
    \end{subfigure}
    \caption{Performance comparison of selection schemes under varying system parameters.}
    \label{fig:combined}
\end{figure}

\subsection{DVQC Incentive Optimization Performance}
 In this subsection, we assess the performance of the proposed ADMM-based incentive optimization scheme for DVQC. We compare our   method with three benchmarks: relaxed upper bound, greedy, and fixed selection, where the relaxed upper bound allows fractional choices in [0,1] to provide an ideal performance ceiling, the greedy method selects processors by highest utility-to-qubit ratio under budget and coverage constraints, and the fixed strategy chooses processors solely by qubit counts, ignoring utility and cost.





Fig.~\ref{fig:energy4schemes} shows the convergence of the proposed ADMM-based incentive optimization versus three baselines: relaxed upper bound, greedy, and fixed selection. The relaxed upper bound serves as an ideal benchmark, achieving the highest utility. ADMM quickly approaches this benchmark within five iterations, demonstrating effective handling of payment and qubit constraints. The greedy method shows only moderate utility and rapidly plateaus because it makes straightforward selections based on a simple utility-to-qubit ratio without iterative refinement, while the fixed scheme yields the lowest, non-improving utility. These results highlight the superior convergence and solution quality of the ADMM approach compared to simpler alternatives. \textcolor{black}{Specifically, the rapid convergence of ADMM is attributed to its alternating optimization strategy, which iteratively refines the processor selection while enforcing the payment budget and qubit coverage constraints through the augmented Lagrangian formulation. Unlike the greedy and fixed schemes, ADMM continuously adjusts the selection variables to maximize system utility. Consequently, the proposed algorithm converges to a solution that closely approaches the relaxed upper bound while maintaining feasibility under the practical constraints of the DVQC incentive framework.}


Fig.~\ref{fig:9} compares the system utility of the four schemes as the payment budget increases. \textcolor{black}{As the payment budget increases, the system utility improves for the relaxed upper bound, ADMM, and greedy schemes with ADMM consistently remaining close to the relaxed upper bound across the entire budget range. This trend is expected because a larger payment budget allows the aggregator to compensate and recruit a more suitable set of processors, thereby enlarging the feasible solution space for maximizing the system utility while satisfying the required qubit coverage. The proposed ADMM algorithm is able to exploit the additional budget more effectively than the greedy and fixed schemes, maintaining performance close to the relaxed upper bound. The greedy scheme achieves only moderate utility improvement because it makes processor selection based on the utility-to-qubit ratio without optimizing the overall system utility. The fixed scheme exhibits constant utility since it selects processors solely according to their qubit capacities, without adapting to the available payment budget.} Fig.~\ref{fig:10} illustrates the impact of the required qubit coverage $Q$ on the system utility. \textcolor{black}{As $Q$ increases, the system utility decreases across all schemes because satisfying the higher qubit coverage requirement becomes more challenging under the same payment budget. Consequently, the aggregator may need to select additional processors or processors with higher energy costs to satisfy the increased qubit requirement, resulting in a lower system utility. The relaxed upper bound provides the highest utility as an ideal benchmark, while ADMM closely matches its performance by effectively balancing the payment budget and qubit coverage constraints. In contrast, the greedy and fixed schemes exhibit a more pronounced utility degradation since their simpler selection strategies can not efficiently trade off processor energy costs against qubit contributions. These results demonstrate that the proposed ADMM framework maintains high system utility even under increasing resource requirements by optimizing processor selection with respect to both incentive and qubit constraints.}

\section{Conclusions} \label{sectionV}
\textcolor{black}{This paper has proposed a two-stage incentive mechanism design for DVQC over wireless networks to enable scalable execution of large PQCs through parallel subcircuit processing. In the first stage, we have addressed the processor-side energy minimization problem by jointly optimizing subcircuit parameters, number of noise measurements, and transmit power, and have solved it using an efficient combination of BCD and SCA. In the second stage, the aggregator has leveraged the optimized energy costs from the first stage to solve a system utility maximization problem, selecting the optimal set of processors. This NP-hard, non-convex combinatorial selection problem has been effectively addressed using the ADMM framework.  Extensive numerical results have demonstrated that our DVQC framework's performance matches centralized quantum computing, while significantly outperforming baseline methods in both energy efficiency and system utility. Future work will focus on extending the framework to incorporate more realistic quantum hardware noise models, and developing adaptive incentive strategies for large-scale distributed quantum computing systems.}

\appendices

\section{Convexification of \eqref{eqn:26a'}} \label{appendixA}

We re-write \eqref{eqn:26a'} as $\alpha \beta \geq s_{k}^{(t)}$. 
\begin{align}
   &\alpha \beta \geq s_{k}^{(t)}
   \Leftrightarrow \frac{1}{4} (\alpha+\beta)^2 - s_{k}^{(t)} \geq  \frac{1}{4} (\alpha-\beta)^2. \label{eqn:27'}
\end{align}
Although the right-hand side of \eqref{eqn:27'} is convex, the left-hand side is still non-convex. Using the first-order Taylor expansion, we approximate $(\alpha+\beta)^2$ as,
\begin{align}
    (\alpha+\beta)^2 \geq (\alpha_{i}+\beta_{i})^2 + 2 (\alpha_{i}+\beta_{i}) (\alpha+\beta-\alpha_{i}-\beta_{i}). \label{eqn:28'}
\end{align}
Putting \eqref{eqn:28'} into \eqref{eqn:27'}, we arrive at
\begin{align}
    \frac{1}{4} \bigg[(\alpha_{i}+\beta_{i})^2 &+ 2 (\alpha_{i}+\beta_{i})(\alpha+\beta-\alpha_{i}-\beta_{i})\bigg] \nonumber \\
    &- s_{k}^{(t)} \geq \frac{1}{4} (\alpha-\beta)^2. \label{eqn:29'}
\end{align}

\section{Update of $x_{k}$} \label{appendixB} 

Fixing $z_{k}^{(i)}$ and $\lambda_{k}^{(i)}$, variable $x_{k}$ is updated by minimizing the augmented Lagrangian over the box constraint $x_{k} \in [0, 1]$ as
\begin{align}
    x_{k}^{(i+1)} &= \arg \min_{x_{k} \in [0, 1]} \bigg\{- \bigg(m_{k} - E_{k}\bigg) x_{k} \nonumber \\
    &+ \frac{\rho}{2} {\bigg(x_{k} - z_{k}^{(i)} + \lambda_{k}^{(i)}\bigg)}^{2}\bigg\}, \quad \forall{k \in \mathcal{K}}.
\end{align}
For minimizing, we now take its derivative and set it to zero as
$
    - \bigg(m_{k} - E_{k}\bigg) x_{k} + \rho  \bigg(x_{k} - z_{k}^{(i)} + \lambda_{k}^{(i)}\bigg) = 0, \forall{k \in \mathcal{K}}.
$
Solving it for $x_{k}$, we obtain
$
    x_{k} = z_{k}^{(i)} - \lambda_{k}^{(i)} + \frac{m_{k} - E_{k}}{\rho}, \quad \forall{k \in \mathcal{K}}.
$
Projecting onto $[0, 1]$, we arrive at
\begin{align}
    x_{k}^{(i+1)} = \min \bigg(1, \max \bigg(0, z_{k}^{(i)} - \lambda_{k}^{(i)} &+ \frac{m_{k} - E_{k}}{\rho}\bigg)\bigg), \nonumber \\ 
    &\forall{k \in \mathcal{K}}.
\end{align}

\bibliographystyle{IEEEtran}
\bibliography{Main-Bibliography}

\begin{IEEEbiography}[{\includegraphics[width=1in,height=1.25in,clip,keepaspectratio]{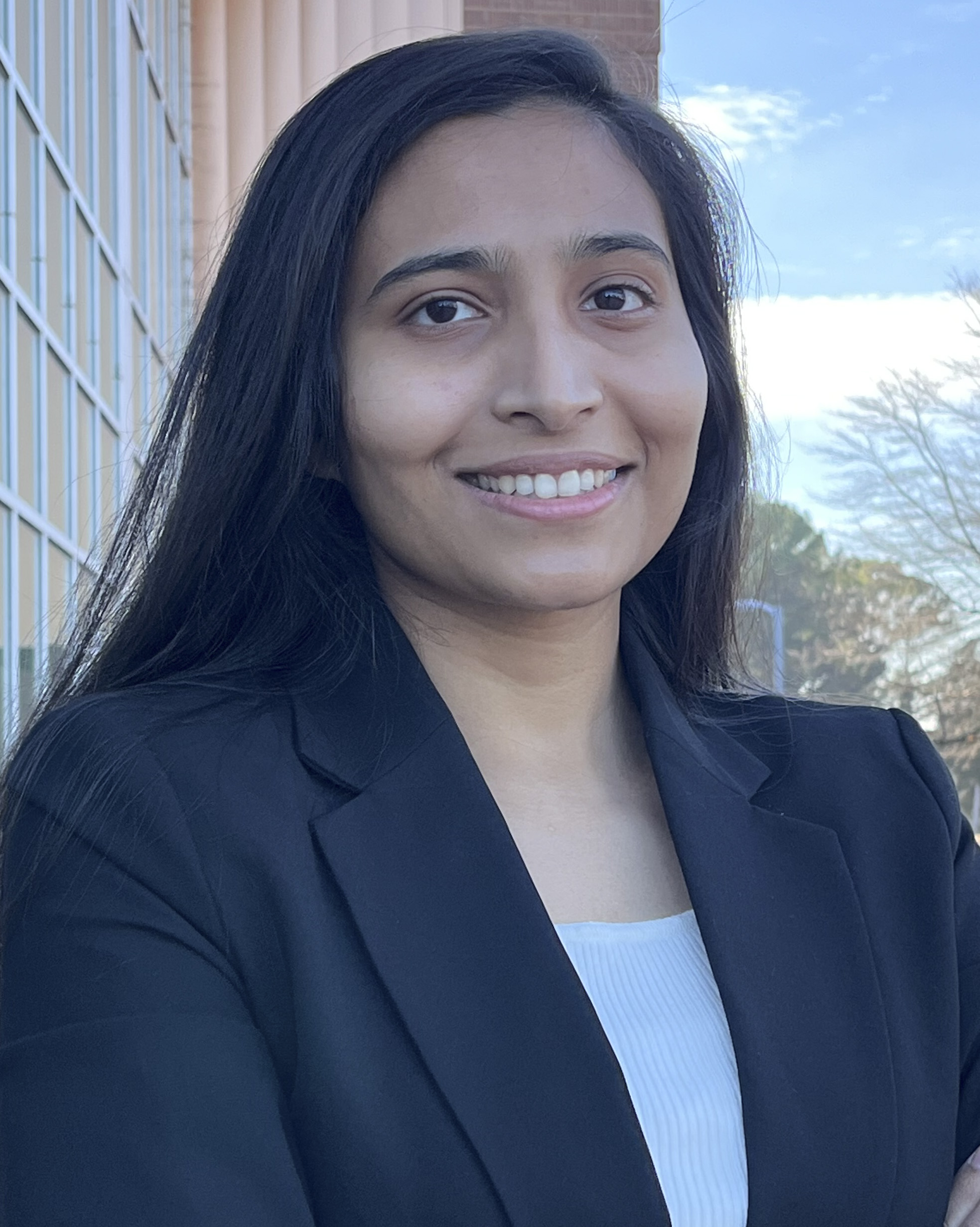}}]{Shaba Shaon (Graduate Student Member, IEEE)} is currently pursuing the Ph.D. degree in Computer Engineering with the Department of Electrical and Computer Engineering at The University of Alabama in Huntsville, Huntsville, AL, USA. She is a member of the Networking, Intelligence, and Security Research Lab at the same institution, where her research focuses on developing intelligent, efficient, and scalable computational frameworks for next-generation wireless networks and quantum-enabled distributed systems. Her research interests include federated learning, wireless networking, distributed quantum computing, distributed quantum sensing, and classical and quantum optimization algorithms.
\end{IEEEbiography}

\begin{IEEEbiography}[{\includegraphics[width=1in,height=1.25in,clip,keepaspectratio]{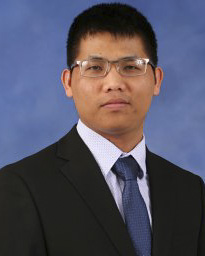}}]{Dinh C. Nguyen (Member, IEEE)} is an Assistant Professor at the Department of Electrical and Computer Engineering, The University of Alabama in Huntsville, USA. He worked as a postdoctoral research associate at Purdue University, USA from 2022 to 2023. He obtained the Ph.D. degree in computer science from Deakin University, Australia in 2021. His current research interests include quantum computing, federated machine learning,  wireless networking, and security.   He is an Associate Editor of IEEE Transactions on Network Science and Engineering and ACM Computing Survey Journal.   He received the Best Editor Award from IEEE Open Journal of Communications Society in 2023.
\end{IEEEbiography}

\begin{IEEEbiography}[{\includegraphics[width=1in,height=1.25in,clip,keepaspectratio]{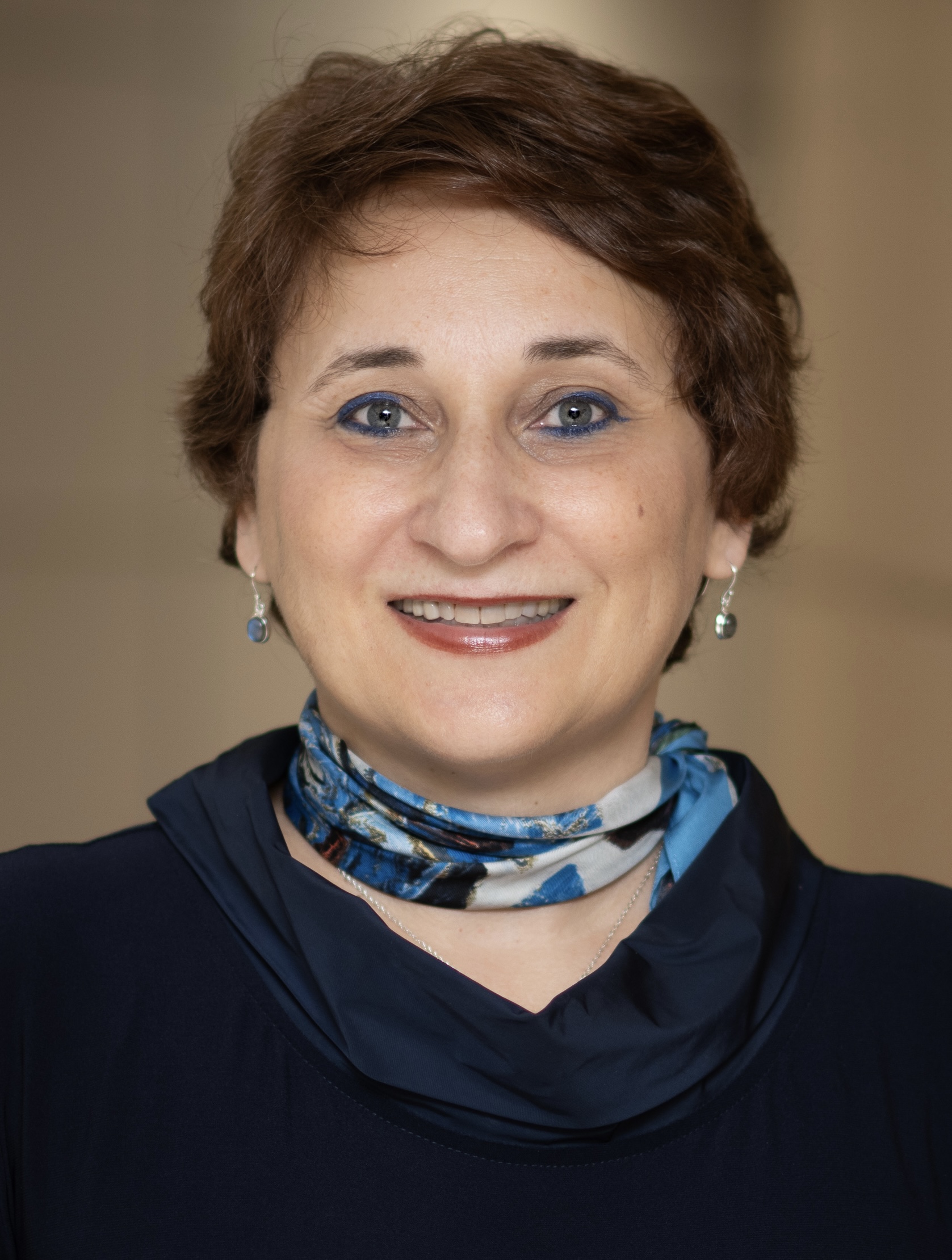}}]{Octavia A. Dobre (Fellow, IEEE)} is a Professor and Tier-1 Canada Research Chair with Memorial University, Canada. Her research spans wireless communications and networking technologies, as well as optical and underwater communications. She was honored with numerous Best Paper Awards, including the IEEE Communications Society Heinrich Hertz Award.
Dr. Dobre served as the VP Publications of the IEEE Communications Society, the founding Editor-in-Chief of the IEEE Open Journal of the Communications Society, and the Editor-in-Chief of the IEEE Communications Letters. Currently, she is Editor-in-Chief of the IEEE Wireless Communications Magazine.
Dr. Dobre is a Clarivate Highly Cited Researcher, a Fellow of the Royal Society of Canada, the Canadian Academy of Engineering, and the Engineering Institute of Canada. She is also an elected member of the European Academy of Sciences and Arts and the Academia Europaea.

\end{IEEEbiography}

\end{document}